\documentclass[a4paper,11pt]{article}
\pdfoutput=1
\usepackage{jcappub}
\usepackage{amsmath}
\usepackage[T1]{fontenc}
\usepackage{bm}
\usepackage{todonotes}
\usepackage{ulem}

\title{\boldmath Thermalisation of sterile neutrinos in the early Universe in the 3+1 scheme with full mixing matrix}

\author[a]{S.\ Gariazzo,}
\emailAdd{gariazzo@ific.uv.es}

\author[b]{P.F.\ de Salas,}
\emailAdd{pablo.fernandez@fysik.su.se}

\author[a]{S.\ Pastor\,}
\emailAdd{pastor@ific.uv.es}

\affiliation[a]{Instituto de F{\'\i}sica Corpuscular  (CSIC-Universitat de Val{\`e}ncia), Valencia, Spain}
\affiliation[b]{The Oskar Klein Centre for Cosmoparticle Physics,
Department of Physics, Stockholm University, SE-106 91 Stockholm, Sweden}

\abstract{
In the framework of a 3+1 scheme with an additional inert state, 
we consider the thermalisation of sterile neutrinos in the early Universe taking into account the full $4\times4$ mixing matrix.
The evolution of the neutrino energy distributions is found solving the momentum-dependent kinetic equations with full diagonal collision terms, 
as in previous analyses of flavour neutrino decoupling in the standard case.
The degree of thermalisation of the sterile state is shown in terms of the effective number of neutrinos, \Neff, and its dependence on the 
three additional mixing angles ($\theta_{14}$, $\theta_{24}$, $\theta_{34}$) and on the squared mass difference \dmsq{41}
is discussed. Our results are relevant for fixing the contribution of a fourth light neutrino species to the cosmological energy density, 
whose value is very well constrained by the final Planck analysis. For the preferred region of active-sterile mixing parameters from
short-baseline neutrino experiments, we find that the fourth state is fully thermalised ($\Neff\simeq 4$).
}

\newcommand{\Neff}{\ensuremath{N_{\rm eff}}}
\newcommand{\DNeff}{\ensuremath{\Delta N_{\rm eff}}}
\newcommand{\dmsq}[1]{\ensuremath{\Delta m^2_{#1}}}
\newcommand{\uasq}[1]{\ensuremath{|U_{#1 4}|^2}}
\newcommand{\e}[1]{\ensuremath{\times10^{#1}}}

\newcommand{\fortepiano}{\texttt{FortEPiaNO}}
\newcommand{\dlsoda}{\texttt{DLSODA}}

\begin{document}

\maketitle

\section{Introduction}

After the initial evidences for neutrino conversion from the detection of solar and atmospheric neutrinos, 
laboratory experiments based at reactors and accelerators were crucial to confirm that the results  were 
explained by the existence of neutrino oscillations.
Nowadays, a vast number of observations nicely fit
in the framework of three-flavour neutrino oscillations, and the last global analyses 
\cite{deSalas:2017kay, Capozzi:2018ubv,Esteban:2018azc} 
provide precise measurements of the mixing parameters, as well as some hints for
the choice of the neutrino mass ordering and improved sensitivity on the CP phase.

There remain, however, a few anomalies found in some short-baseline oscillation expe\-ri\-ments that could indicate the
presence of an additional light neutrino at the eV mass scale that mixes with the ordinary active states
(see e.g.\ the reviews \cite{Abazajian:2012ys,Gariazzo:2015rra,Giunti:2019aiy}).
This problem can be studied assuming a small mixing of this new sterile state with the active neutrinos, the so-called
3+1 scheme\footnote{The cases with more than one sterile state, such as the 3+2 scheme, are disfavoured
by the fact that the larger number of parameters does not guarantee an improvement of the fit with respect to the 3+1 case
\cite{Maltoni:2007zf,Melchiorri:2008gq,Archidiacono:2012ri,Kopp:2013vaa,Gariazzo:2015rra}.},
although recent global analyses of short-baseline data \cite{Gariazzo:2017fdh,Dentler:2018sju,Collin:2016rao}
show that this does not provide an optimal solution due to the severe tension between the anomalies 
in the appearance sector and disappearance measurements.
It is expected that new data, from both running 
and forthcoming neutrino experiments, will shed light on the causes of this tension and eventually provide a definitive solution to this puzzle.
In the meantime, it is interesting to explore the implications of this kind of active-sterile oscillations in astrophysical and cosmological scenarios.

Concerning cosmology, a well-known consequence of active-sterile oscillations would be the production of the new states in the early Universe.
If oscillations become effective before neutrino decoupling, the sterile species will appear via mixing while the active states keep an equilibrium energy distribution.
The degree of thermalisation depends on the specific values of the neutrino parameters and fixes the contribution of all neutrino states to the 
cos\-mo\-lo\-gical energy density of relativistic particles, usually parametrised by the effective number of neutrinos (\Neff).
A fully thermalised fourth neutrino state would lead to
a value $\Neff \simeq 4$, which is disfavoured according to the analysis of the full-mission data from the Planck satellite \cite{Aghanim:2018eyx} on the anisotropies of the cosmic microwave background (CMB).
Combined with other cosmological and astrophysical measurements, the allowed range can be as restricted as $\Neff = 2.99^{+0.34}_{-0.33}$ (95\% confidence region).
Thus, it is important to be able to perform a proper calculation, as precise as possible, of the values of \Neff\ for each choice of the mixing parameters describing 
the active-sterile neutrino oscillations.

The cosmological evolution of the active-sterile neutrino system in the early Universe is a complex problem due to the simultaneous presence of oscillations in a changing medium 
and effective weak interactions.
It has been studied in a large number of previous papers, where the corresponding Boltzmann kinetic equations were solved including different approximations.
The first works simplified the task considering mixing with only one neutrino state and that the neutrinos were well described by a single, average neutrino momentum 
(see e.g.\ \cite{Barbieri:1989ti,Kainulainen:1990ds,Barbieri:1990vx,Enqvist:1990ad,Enqvist:1991qj} and the review \cite{Dolgov:2002wy} for a complete list of early references).
Later studies have dealt with more realistic cases, including the dependence on neutrino momenta and/or mixing of the sterile state with two or more active neutrinos.
 
In principle, solving the Boltzmann equations for unequal neutrino momenta is mandatory, because both oscillations and collisions present a (different) dependence on the neutrino energy.
Moreover, these processes can lead to non-thermal distortions in the neutrino energy distributions that would be only found in multi-momentum calculations, with the least 
number of assumptions as possible.
This has been achieved in the case of three active neutrinos, where the standard value $\Neff=3.045$ was found \cite{deSalas:2016ztq}, but the 
computational problem is more demanding when active-sterile mixing is included.
If the mass difference with the mostly ste\-rile state is very small ($\Delta m^2 \leq 10^{-7}$ eV$^2$),
collisions can be neglected and the evolution of the neutrino spectra can be calculated very precisely 
\cite{Kirilova:1997sv,Kirilova:1999xj,Kirilova:2006wh}, but there is no enhancement of \Neff.

In the recent years, several authors (see e.g.\ \cite{Hannestad:2012ky,Hannestad:2015tea}) have presented multi-momentum calculations of 
active-sterile oscillations in the early Universe in the approximation of only one active and one sterile neutrino species (1+1 scenario), in some 
cases including a potentially large lepton asymmetry.
In particular, the first results in the 1+1 case ($\nu_e$-$\nu_s$) with full collision integrals 
were shown in \cite{Hannestad:2015tea} and the values found for \Neff\ were compared with those obtained with various approximations.
The quantum kinetic equations were solved with a modified version of the \texttt{LASAGNA} code \cite{Hannestad:2013wwj}, enforcing a zero lepton asymmetry.
The same code has been used in later works \cite{Bridle:2016isd,Guzowski:2017ryo,Knee:2018rvj,Berryman:2019nvr} to convert the active-sterile mixing parameters 
into two other quantities relevant for cosmology ($\Neff$ and the effective sterile neutrino mass $m^s_{\rm eff}$), in order to obtain bounds from Planck data 
and from current and future neutrino oscillation experiments in the framework of the two-neutrino approximation (either $\nu_e$-$\nu_s$ or
$\nu_\mu$-$\nu_s$ mixing).

On the other hand, a precise calculation of 3+1 active-sterile oscillations in cosmology must include the unavoidable presence of mixing among active neutrinos
(early simplified analyses include \cite{Dolgov:2003sg,Cirelli:2004cz}), i.e.\ the full four-neutrino mixing matrix with up to six different angles: three exclusive 
of the active sector ($\theta_{12}$, $\theta_{13}$, $\theta_{23}$) and three related to the mixing with the sterile state
($\theta_{14}$, $\theta_{24}$, $\theta_{34}$).
More recent multi-angle studies \cite{Mirizzi:2012we,Mirizzi:2013gnd} have been performed within the averaged-momentum approximation.
In particular, the authors of \cite{Mirizzi:2013gnd} have shown how the cosmological constraints change if two active-sterile mixing angles are considered.
A first step beyond the averaged-momentum and single-mixing approximations was taken in \cite{Saviano:2013ktj}.
This study considers  
a multi-momentum and multi-flavour calculation of the kinetic equations of the active-sterile system in the 2+1 scenario, with mixing parameters 
inspired by the short-baseline neutrino anomalies and in the presence of primordial neutrino asymmetries, where the production of the sterile state can be suppressed.
However, to our notice there is no code available to compute the neutrino evolution in the full 3+1 case with momentum dependence.

Prompted by the current precision on the determination of the effective number of neutrinos by Planck data 
and taking advantage of our previous experience on multi-momentum calculations in the standard three-neutrino case \cite{Mangano:2005cc,deSalas:2016ztq},
in this paper we present an analysis of the sterile neutrino thermalisation in the early Universe in the 3+1 scenario, based on the numerical solution of the kinetic equations 
with full collision terms and the complete $4\times 4$ mixing matrix.

The rest of this paper is organised as follows.
In section \ref{sec:eqs} we present the equations for active-sterile oscillations in the 3+1 scenario, which
are solved with a new numerical code.
We describe in section \ref{sec:results} the main results concerning the evolution of cosmological neutrinos, with emphasis on
the final values of the effective number of neutrinos and the dependence on the mixing parameters.
Our main conclusions are summarised in \ref{sec:conc}, while the two appendices are devoted to technical details on the collision terms of the kinetic equations and
to a description of our computational code.

\section{Active-sterile oscillations in the early Universe}
\label{sec:eqs}

In order to calculate the degree of thermalisation of light sterile neutrinos via oscillations, we need to solve the
corresponding set of quantum kinetic equations in an expanding Universe. For the relevant cosmological temperatures, from
a few tens or hundreds of MeV down to a few keV, both neutrino oscillations and interactions are important. Therefore,
in order to take into account all effects we consider the evolution of a $4\times4$ neutrino density matrix, defined as
\begin{equation}
\varrho(p, t)
=
\left(
\begin{array}{cccc}
\varrho_{ee}&\varrho_{e\mu}&\varrho_{e\tau}&\varrho_{es}\\
\varrho_{\mu e}&\varrho_{\mu\mu}&\varrho_{\mu\tau}&\varrho_{\mu s}\\
\varrho_{\tau e}&\varrho_{\tau\mu}&\varrho_{\tau\tau}&\varrho_{\tau s}\\
\varrho_{se}&\varrho_{s\mu}&\varrho_{s\tau}&\varrho_{ss}\\
\end{array}
\right)\,.
\label{rho_4x4}
\end{equation}
Each term of $\varrho$ depends on the neutrino momentum $p$ and evolves with time $t$. Since we 
consider no lepton asymmetry, the density matrices for neutrinos and antineutrinos are the same.
The (real) diagonal terms $\varrho_{\alpha \alpha}$ contain the momentum distribution function of the neutrino flavour $\alpha$,
while the off-diagonal terms are complex and take values different from zero only in presence of mixing.

The evolution of the density matrix for a given momentum $p$ is dictated by the Boltzmann equations 
\cite{Sigl:1992fn,deSalas:2016ztq,Mirizzi:2012we,Saviano:2013ktj}, which in compact form read as
\begin{equation}\label{eq:drho_dx_no-comoving_4x4}
\left( \partial_t - H p\, \partial_p \right) \varrho(t) =
-i \left[
	\left(
	\frac{1}{2p}\mathbb{M}_{\rm F}
	- \frac{8\sqrt{2}G_{\rm F} p}{3} \left( \frac{\mathbb{E}_l}{m_W^2} +  \frac{\mathbb{E}_\nu}{m_Z^2}\right)
	\right)
	, \varrho (t) 
\right]
+ \mathcal{I}\left[\varrho(t) \right],
\end{equation}
where $m_W$ and $m_Z$ are the $W$ and $Z$ boson masses, $G_{\rm F}$ is the Fermi constant,
and $H$ is the Hubble expansion rate. The neutrino mass matrix in the flavour basis is
$\mathbb{M}_{\rm F}=U\mathbb{M}U^\dagger$,
rotated from the diagonal mass matrix $\mathbb{M}=\text{diag}(m_1^2,\ldots,m_4^2)$ using the $4\times 4$ mixing matrix $U$.
The other two matrices in the commutator,
\begin{equation}\label{eq:matterpotentials}
\mathbb{E}_\ell=\text{diag}(\rho_e, \rho_\mu, 0, 0)\,,
\qquad
\mathbb{E}_\nu=S_a\left(\int {\rm d}y y^3\varrho\right) S_a\,,
\quad\mbox{with }
S_a=\text{diag}(1,1,1,0)\,,
\end{equation}
define the matter potential for neutrino oscillations, accounting for the energy densities of charged leptons ($\rho_\ell = \rho_{\ell^-} + \rho_{\ell^+}$, where $\ell=e,\mu$ since the $\tau$ density is negligible for the relevant temperatures) and neutrinos. The latter term includes diagonal and off-diagonal components from
active neutrinos \cite{Sigl:1992fn}. Finally, 
neutrino non-forward interactions are encoded in the collision term $\mathcal{I(\varrho)}$, described in detail in appendix~\ref{sec:collint}.
In Eq.~\eqref{eq:drho_dx_no-comoving_4x4} we therefore take into account
neutrino oscillations in matter (first term in the right-hand side),
neutrino interactions with the particles of the cosmic plasma (last term in the right-hand side),
and the expansion of the Universe (second term in the left-hand side).

There are different ways that can be adopted to numerically solve the Boltzmann equations. Here we choose 
to discretise the momentum of the incoming neutrino and use, as in previous studies, 
the comoving variables $x\equiv m_e\, a$, $y\equiv p\, a$ and $z\equiv T_\gamma\, a$,
with the electron mass $m_e$ chosen as an arbitrary mass scale
and $T_\gamma$ the photon temperature.
The scale factor $a=1/T$ that we use in order to compute comoving quantities is normalised according to
$T^0=T_\gamma^0=1$ at early times, where the temperature $T$ is only initially equal to the photon temperature,
but later does not represent the real temperature of any of the particles in the plasma.
Written in terms of these variables, the evolution of the density matrix is \cite{deSalas:2016ztq,Mirizzi:2012we,Saviano:2013ktj}~%
\footnote{We omit the bar notation adopted for example in \cite{deSalas:2016ztq}
to denote the quantities expressed in terms of the comoving variables.}
\begin{equation}\label{eq:drho_dx}
\frac{{\rm d}\varrho(y,x)}{{\rm d}x}
=
\sqrt{\frac{3 m^2_{\rm Pl}}{8\pi\rho}}
\left\{
    -i \frac{x^2}{m_e^3}
    \left[
        \frac{\mathbb{M}_{\rm F}}{2y}
        -
        \frac{8\sqrt{2}G_{\rm F} y m_e^6}{3x^6}
        \left(
            \frac{\mathbb{E}_\ell}{m_W^2}
            +
            \frac{\mathbb{E}_\nu}{m_Z^2}
        \right),
    \varrho
    \right]
    +\frac{m_e^3}{x^4}\mathcal{I(\varrho)}
\right\}
\end{equation}
where $m_{\rm Pl}$ is the Planck mass
and $\rho$ is the total comoving energy density, i.e.\ the physical energy density multiplied by a factor $a^4$.

For the neutrino mixing matrix,
which in principle is parametrisation-independent but can be conveniently written in terms of mixing angles,
we use the convention presented in Eq.\ (12) of \cite{Gariazzo:2015rra},
but extended to the full $4\times4$ unitary $U$,
without considering any of the CP violating phases, which are all fixed to zero.
Thus, the mixing matrix is
\begin{equation}\label{eq:mixing_matrix}
U=R^{34} R^{24} R^{14} R^{23} R^{13} R^{12},
\end{equation}
where each $R^{ij}$ is a real rotation matrix described by the angle $\theta_{ij}$,
containing $\cos\theta_{ij}$ in the diagonal elements $ii$ and $jj$,
1 in the remaining diagonal elements,
$\sin\theta_{ij}$ ($-\sin\theta_{ij}$) in the off-diagonal element $ij$ ($ji$)
and zero otherwise:
\begin{equation}
\label{eq:rotationmatrix}
[R^{ij}]_{rs}=
\delta_{rs}
+
(\cos\theta_{ij}-1)(\delta_{ri}\delta_{si}+\delta_{rj}\delta_{sj})
+
\sin\theta_{ij}(\delta_{ri}\delta_{sj}-\delta_{rj}\delta_{si})\,.
\end{equation}
Our complete case can therefore be described using six angles, of which
$\theta_{12}$, $\theta_{13}$ and $\theta_{23}$
characterise the active neutrino mixing and can be obtained by standard three-neutrino global fits.
Specifically, we use the best-fit values from \cite{deSalas:2017kay},
focusing mainly on the normal ordering of active neutrino masses,
which is currently favoured (see e.g.~\cite{Gariazzo:2018pei,deSalas:2018bym}).
In addition, we describe the active-sterile mixing using
the parametrisation-independent entries of the fourth column of the mixing matrix,
which can be expressed
as functions of the new angles $\theta_{14}$, $\theta_{24}$ and $\theta_{34}$
with the parametrisation in Eq.~\eqref{eq:mixing_matrix}:
\begin{eqnarray}
\uasq{e} & = & \sin^2\theta_{14},\nonumber \\
\uasq{\mu} & = & \cos^2\theta_{14}\sin^2\theta_{24}, \nonumber \\
\uasq{\tau} & = & \cos^2\theta_{14}\cos^2\theta_{24}\sin^2\theta_{34}, \nonumber \\
\uasq{s} & = & \cos^2\theta_{14}\cos^2\theta_{24}\cos^2\theta_{34}.
\label{eq:mixing_fourth_col}
\end{eqnarray}
The list of mixing parameters includes also the two standard mass splittings
\dmsq{21}, \dmsq{31} and the new \dmsq{41},
which we use to define the diagonal mass matrix $\mathbb{M}$.
For the active-sterile mixing parameters,
we consider as benchmark the mass splitting $\dmsq{41}=1.29$~eV$^2$
and the mixing matrix element $\uasq{e}\simeq0.012$,
as currently favoured by the fit of electron antineutrino disappearance data
from the DANSS \cite{Alekseev:2018efk,Egorov:2018nu}
and NEOS \cite{Ko:2016owz}
experiments in the context of the 3+1 scenario \cite{Gariazzo:2018mwd,Dentler:2017tkw,Dentler:2018sju}.
Since flavour oscillations are blind to global phases, we substract the mass of the lightest neutrino, which does not enter the calculations as long as neutrinos are ultra-relativistic.

In addition to the Boltzmann equations we need to solve the continuity equation for the total energy density of radiation, which in terms of non-comoving variables is
\begin{equation}\label{eq:continuity}
\frac{\mathrm{d} \rho}{\mathrm{d}t} = -3 H \left( \rho + P \right),
\end{equation}
where $P$ is the pressure. This last equation provides the evolution of the comoving photon temperature $z$ as a function of the comoving variable $x$,
and can be conveniently written in terms of $r=x/z$ as in \cite{Mangano:2001iu}.
We also take into account the finite temperature QED corrections,
since the frequent interactions that keep particles in equilibrium in the cosmic plasma
also contribute as an effective correction to their masses.
As a result of these corrections, the evolution of $z$ and neutrino interactions with electrons and positrons are modified.
We include these modifications as explained in \cite{Mangano:2001iu,Heckler:1994tv,Fornengo:1997wa}.

When taking into account the electromagnetic corrections
and the contribution coming from the relevant charged leptons,
Eq.~\eqref{eq:continuity} can be rewritten to obtain
the evolution of $z$ as a function of $x$~\cite{Mangano:2001iu} as
\begin{equation}\label{eq:dz_dx_4x4}
\frac{\mathrm{d}z}{\mathrm{d}x}
=
\cfrac{
{\displaystyle \sum_{\ell=e,\mu}}
\left[
\cfrac{r_\ell^2}{r} J(r_\ell)
\right]
+ G_1(r)
- \cfrac{1}{2\pi^2z^3}
    {\displaystyle \int_0^\infty {\rm d}y\,y^3\sum_{\alpha=e}^s\cfrac{\mathrm{d}\varrho_{\alpha \alpha}}{\mathrm{d}x}}
}{
{\displaystyle \sum_{\ell=e,\mu}}
\left[
r^2_\ell J(r_\ell)
+ Y(r_\ell)
\right]
+ G_2(r)
+ \cfrac{2\pi^2}{15}
}\,,
\end{equation}
where the $\mathrm{d}\varrho_{\alpha \alpha}/\mathrm{d}x$ are obtained from Eq.~\eqref{eq:drho_dx}
and $r_\ell=m_\ell/m_e\,r$.
The expressions for the $J$, $Y$, $G_1$ and $G_2$ functions are written in Eqs.~(18)--(22) of \cite{Mangano:2001iu}
and can be found in appendix~\ref{sec:fortepiano}.

Sterile neutrinos are produced via oscillations with the active species, which
become effective when the oscillation frequencies overcome the interaction rate.
If $\dmsq{41} \sim 1\,\mathrm{eV}^2$, active-sterile oscillations are faster than the 
standard ones, and therefore they may begin to populate the mostly sterile state
when the presence of muons in the cosmic plasma is still important.
For this reason we include not only electrons, but also muons in Eq.~\eqref{eq:dz_dx_4x4}
and in the matter potentials of Eq.~\eqref{eq:drho_dx}.
The presence of pions and other hadrons can be neglected, however,
since they interact evenly with active neutrinos and
therefore their effect on oscillations is small when compared with that of charged leptons;
furthermore, from the point of view of the evolution of $z$, they act as an overall shift at late times.

Although it is not necessary for our calculations,
it is interesting to estimate the effective temperature of neutrinos.
This can be used in order to approximate the final neutrino distribution as a Fermi-Dirac,
with an error that is typically at most of the order of one percent for active neutrinos.
For the sterile neutrino, we will comment later that the final momentum distribution function
depends on whether the thermalisation is complete or not.
An estimate of the comoving neutrino temperature $w\equiv T_\nu\, a$
can be obtained from Eq.~\eqref{eq:dz_dx_4x4}, 
considering the process of neutrino decoupling but neglecting electron-positron annihilations assuming they were always relativistic.
This implies that the effective neutrino temperature will follow the one of photons until
the $e^+e^-$ annihilations start to transfer energy to the photon fluid,
after neutrino decoupling,
and will remain constant at later times.

In this paper we use the above equations to compute the evolution of neutrino flavours in cosmology
by means of a new code named \fortepiano\
(\texttt{FORTran-Evolved PrimordIAl Neutrino Oscillations}),
which is described in details appendix~\ref{sec:fortepiano}.
The code considers a grid of neutrino momenta,
distributed according to the Gauss-Laguerre quadrature method, 
to parametrise the density matrix,
and evolves the differential equations presented above over a wide range of comoving temperatures,
through the adaptive solver for stiff problems \dlsoda. 
\fortepiano\ can be used to compute oscillations with up to six neutrinos in a flexible way.
As a comparison, when we only consider the standard case of three active neutrinos,
we obtain $\Neff=3.044$, with very small variations due to the technical settings.
Instead, within the 1+1 neutrino approximation,
the code leads to results in reasonable agreement with \texttt{LASAGNA} \cite{Hannestad:2013wwj} in the relevant range.

\section{Thermalisation of the light sterile neutrinos}
\label{sec:results}

After discussing the evolution of the cosmological energy and entropy densities in the range of relevant temperatures, in 
this section we describe our main results concerning the production and thermalisation of sterile neutrinos via oscillations
and later we report how they are affected by the choice of the mixing parameters.

\subsection{Energy and entropy conservation}

Although all processes are well known, the analysis of cosmological thermodynamics at the epoch 
when active-sterile oscillations become effective provides a better understanding of how the new
states are thermalised. The chain of relevant processes, considering that they are independent 
(or happen at different scales, as if they were instantaneous from the point of view of the other), involve the following particles in equilibrium:
\begin{subequations}\label{eq:equi-processes}
\begin{align}
\mu^\pm + \overset{(-)}{\nu}_{e,\mu,\tau} + e^\pm + \gamma & \quad \rightarrow \quad g_1 = \frac{57}{4}\,,\label{eq:equi-processes:1-mu}\\
\overset{(-)}{\nu}_{e,\mu,\tau} + e^\pm + \gamma & \quad \rightarrow \quad  g_2 = \frac{43}{4}\,, \label{eq:equi-processes:2-std}\\
\overset{(-)}{\nu}_s + \overset{(-)}{\nu}_{e,\mu,\tau} + e^\pm + \gamma & \quad \rightarrow \quad g_3 = \frac{50}{4}\,, \label{eq:equi-processes:3-sterile}\\
e^\pm + \gamma & \quad \rightarrow \quad g_4 = \frac{22}{4}\,, \label{eq:equi-processes:4-nu-dec}\\
\gamma & \quad \rightarrow \quad g_5 = \frac{8}{4}\,,
\label{eq:equi-processes:5-e}
\end{align}
\end{subequations}
where, as usual, in each case $g$ denotes the degrees of freedom of the relativistic particles involved, assuming that they share the same temperature.
In the first process, from Eq.~\eqref{eq:equi-processes:1-mu} to Eq.~\eqref{eq:equi-processes:2-std},
muons and antimuons annihilate into active neutrinos, $e^\pm$ and photons.
In the next two transitions sterile neutrinos are populated through oscillations (Eq.~\eqref{eq:equi-processes:2-std} to Eq.~\eqref{eq:equi-processes:3-sterile})
and active neutrinos decouple when weak interactions become ineffective (Eq.~\eqref{eq:equi-processes:3-sterile} to Eq.~\eqref{eq:equi-processes:4-nu-dec}), respectively.
Finally, the fourth process, from Eq.~\eqref{eq:equi-processes:4-nu-dec} to Eq.~\eqref{eq:equi-processes:5-e}, represents the electron-positron pair annihilations into photons.
We have neglected the presence of pions and other hadrons before Eq.~\eqref{eq:equi-processes:1-mu} because they act as an overall shift in $z$ evolution, since 
their abundance is even smaller than that of muons. 

\begin{figure}[t]
\centering
\includegraphics[width=0.8\textwidth]{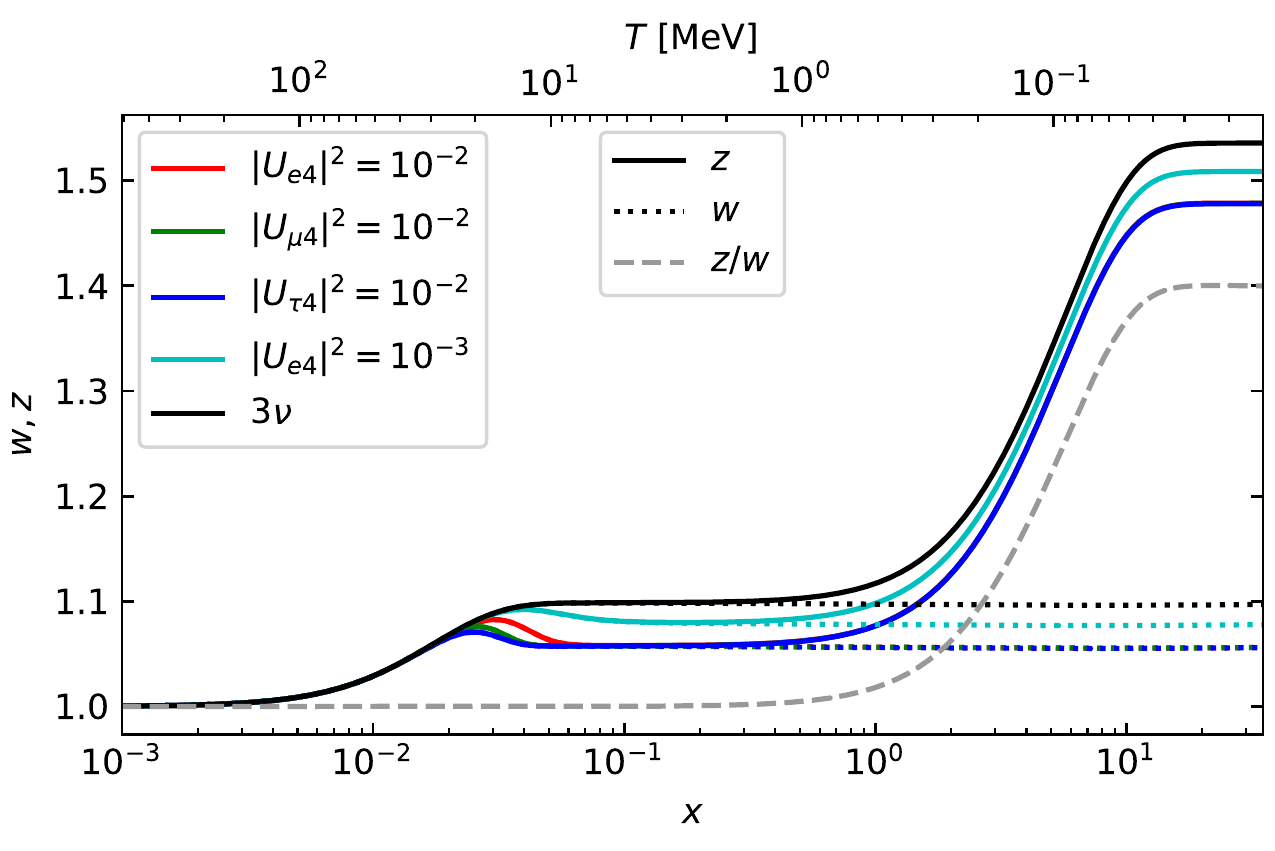}
\caption{\label{fig:z_3p1}
Evolution of $z$ as a function of $x$ for the standard three-neutrino case (black)
compared with some 3+1 cases
(all using $\dmsq{41}=1.29$~eV$^2$ \cite{Gariazzo:2018mwd}, the angles not specified in the legend are set to zero).
One of the 3+1 cases corresponds to $\Neff\simeq3.5$ (cyan), the others to $\Neff\simeq4.05$.
}
\end{figure}

If we now use the comoving temperature $z$ defined in the previous section, for any process
in the early Universe the conservation of the entropy density implies that
\begin{equation}\label{eq:g-entropy-conservation}
g^{\rm s}_{\rm after} \,z^3_{\rm after} = g^{\rm s}_{\rm before} \,z^3_{\rm before}\,,
\end{equation}
while, if the energy density is preserved,
\begin{equation}\label{eq:g-energy-conservation}
g_{\rm after} \,z^4_{\rm after} = g_{\rm before} \,z^4_{\rm before}\,.
\end{equation}
When the temperatures of all the species in thermal equilibrium are the same, $g = g^{\rm s}$.
\begin{figure}[t]
\centering
\includegraphics[width=0.8\textwidth]{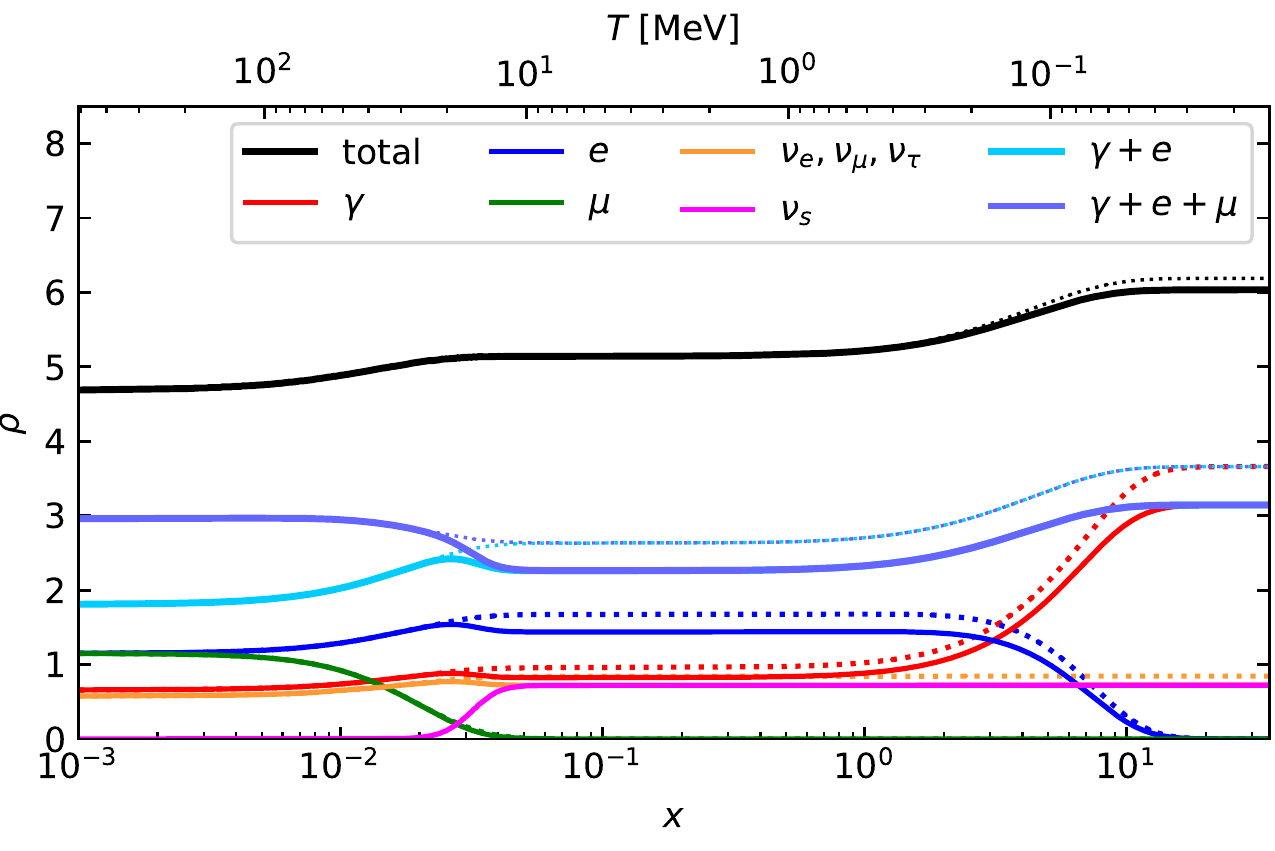}
\includegraphics[width=0.8\textwidth]{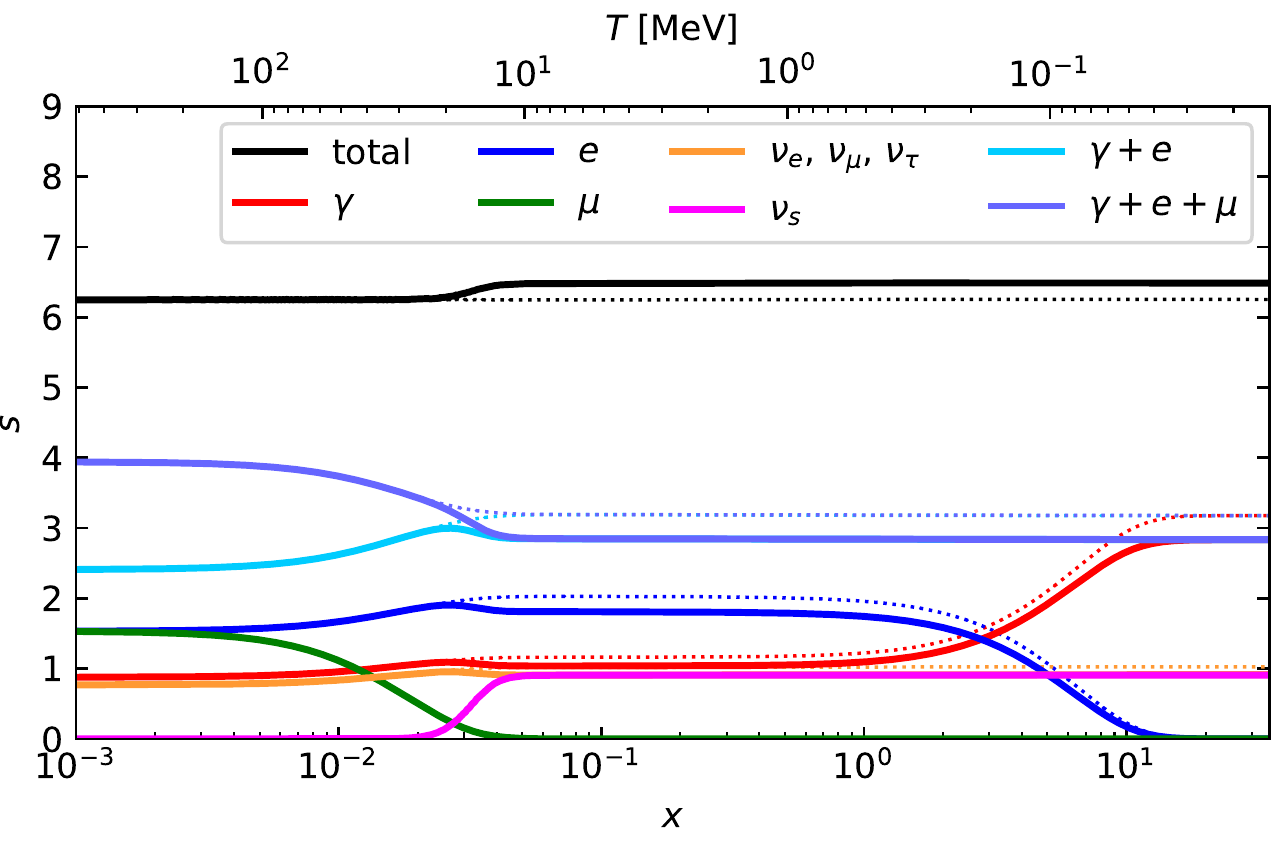}
\caption{\label{fig:energyDensity_entropy}
Evolution of the comoving energy (upper panel) and entropy (lower panel) densities of the different components.
We show the standard three-neutrino case (dotted lines) and a 3+1 case (solid lines)
using $\dmsq{41}=1.29$~eV$^2$,
$\uasq{e}=0.01$ \cite{Gariazzo:2018mwd}, $\uasq{\mu}=\uasq{\tau}=0$,
which gives $\Neff\simeq4.05$.
}
\end{figure}
We show in Fig.\ \ref{fig:energyDensity_entropy} the evolution of the comoving densities $\rho$ (energy) and $s$ (entropy, defined
as $a^3$ times the physical entropy density)
of all relativistic particles, comparing one 3+1 case (solid lines) where
the sterile state is thermalised with the standard case of three active neutrinos (dotted lines).
The corresponding evolution of the comoving temperatures is shown in Fig.\ \ref{fig:z_3p1},
including a 3+1 case where the $\nu_s$ is not fully brought into equilibrium.

Starting at temperatures above 100 MeV,
one can first note that $\mu^\pm$ annihilations increase the energy density of radiation.
During this process the \textit{total} number of degrees of freedom is the same, but
those corresponding to muon and antimuons disappear from the relativistic bath of particles in equilibrium.
As a result, the total entropy density is conserved but the energy density of relativistic particles rises, as can be seen 
in Fig.\ \ref{fig:energyDensity_entropy}. This leads, from Eq.~\eqref{eq:g-entropy-conservation}, to a change in $z$, which increases as shown 
in Fig.~\ref{fig:z_3p1} by an amount
\begin{equation}\label{eq:z-change-muon-ann}
z_2 = \left( \frac{57}{43} \right)^{1/3} z_1\,.
\end{equation}

The second process, the thermalisation of sterile neutrinos via oscillations, enlarges the number of relativistic degrees of freedom.
Since there is no energy injection from annihilating or decaying particles, the total energy density is conserved (Fig.~\ref{fig:energyDensity_entropy}).
However, now there are new degrees of freedom in the game and the entropy density is increased with respect to the three-neutrino case,
as shown in Fig.~\ref{fig:energyDensity_entropy}. If sterile neutrinos are fully thermalised, the comoving temperature of the particles in equilibrium (equivalently $z$)
decreases, as can be seen in Fig.~\ref{fig:z_3p1}, by an amount given by Eq.~\eqref{eq:g-energy-conservation},
\begin{equation}\label{eq:z-change-sterile-production}
z_3 = \left( \frac{43}{50} \right)^{1/4} z_2\,.
\end{equation}

The third process, neutrino decoupling, implies no energy injection, so the energy density is conserved and no new degrees of freedom appear.
The only change that happens is that the neutrino degrees of freedom are now associated to decoupled particles.
This means that, if something happens to the plasma that changes its temperature, the
neutrino comoving temperature $w$ will be constant because it is only an effective parameter to describe their frozen distribution function.
Neutrino decoupling does not change $z$, $\rho$ or $s$, because the degrees of freedom of neutrinos contribute as if they were still coupled to the plasma.
This is shown, respectively, in Fig.~\ref{fig:z_3p1} and in the two panels of Fig.~\ref{fig:energyDensity_entropy}.
Hence,
\begin{equation}\label{eq:z-change-nu-dec}
z_4 = z_3\,.
\end{equation}

The final process, electron-positron pair annihilations, takes place after neutrino decoupling,
increasing the relativistic energy density but conserving the entropy density. Therefore, from Eq.~\eqref{eq:g-entropy-conservation}, 
the change in $z$, which is not felt by the decoupled neutrinos, is given by
\begin{equation}\label{eq:z-change-e-ann}
z_5 = \left( \frac{11}{4} \right)^{1/3} z_4\,.
\end{equation}
This value is the well-known ratio between the photon temperature before and after electron-positron pair annihilations.
Independently of the processes that took place earlier, the ratio of photon and neutrino comoving temperatures always remains the same after $e^\pm$ annihilations.
As shown in Fig.~\ref{fig:z_3p1}, the final $z/w$ is very close to $(11/4)^{1/3}\simeq1.401$.
It is worth commenting that the final energy density appears to be higher in the standard three-neutrino case than in the 3+1 example 
(see the black curves in the upper panel of Fig.\ \ref{fig:energyDensity_entropy}).
This is again a consequence of the conservation of entropy during $e^\pm$ annihilations.
Since the energy density of electrons and positrons is larger when there is no sterile neutrino,
the amount of energy that they can transfer to photons is also increased\footnote{Note that in our normalisation we use
$T_\gamma^0=1$  at very early times, but what we observe today as 
$T_\gamma^{\rm now}$ (one of the best constrained quantities in cosmology)
is the final photon temperature at the end of the evolution.
A proper interpretation of physical quantities in our results (such as the cosmological energy density) 
would therefore require to compute the expected photon temperature $T_\gamma'$
after neutrino decoupling which reproduces $T_\gamma^{\rm now}$ in a given cosmology,
and to renormalise according to $T_\gamma'$ instead of $T_\gamma^0$.
However, a ratio such as \Neff\ is unaffected.}.

This simplified description is valid when the four physical processes follow a sequential order without overlap. In the real cases
some of them coincide to some extent and the computed values of $z$ will differ from that found from 
Eqs.\ \eqref{eq:z-change-muon-ann}--\eqref{eq:z-change-e-ann}.
For instance, the production of sterile neutrinos can take place when $\mu^\pm$ annihilations are still effective, while it is 
well known that active neutrino decoupling is not an instantaneous process:
some relic $e^\pm$ annihilations to neutrinos exist,
leading to the value $\Neff=3.045$ \cite{deSalas:2016ztq}.

\subsection{Effective number of neutrinos}
\label{ssec:neff}

\begin{figure}[t]
\centering
\includegraphics[width=0.8\textwidth]{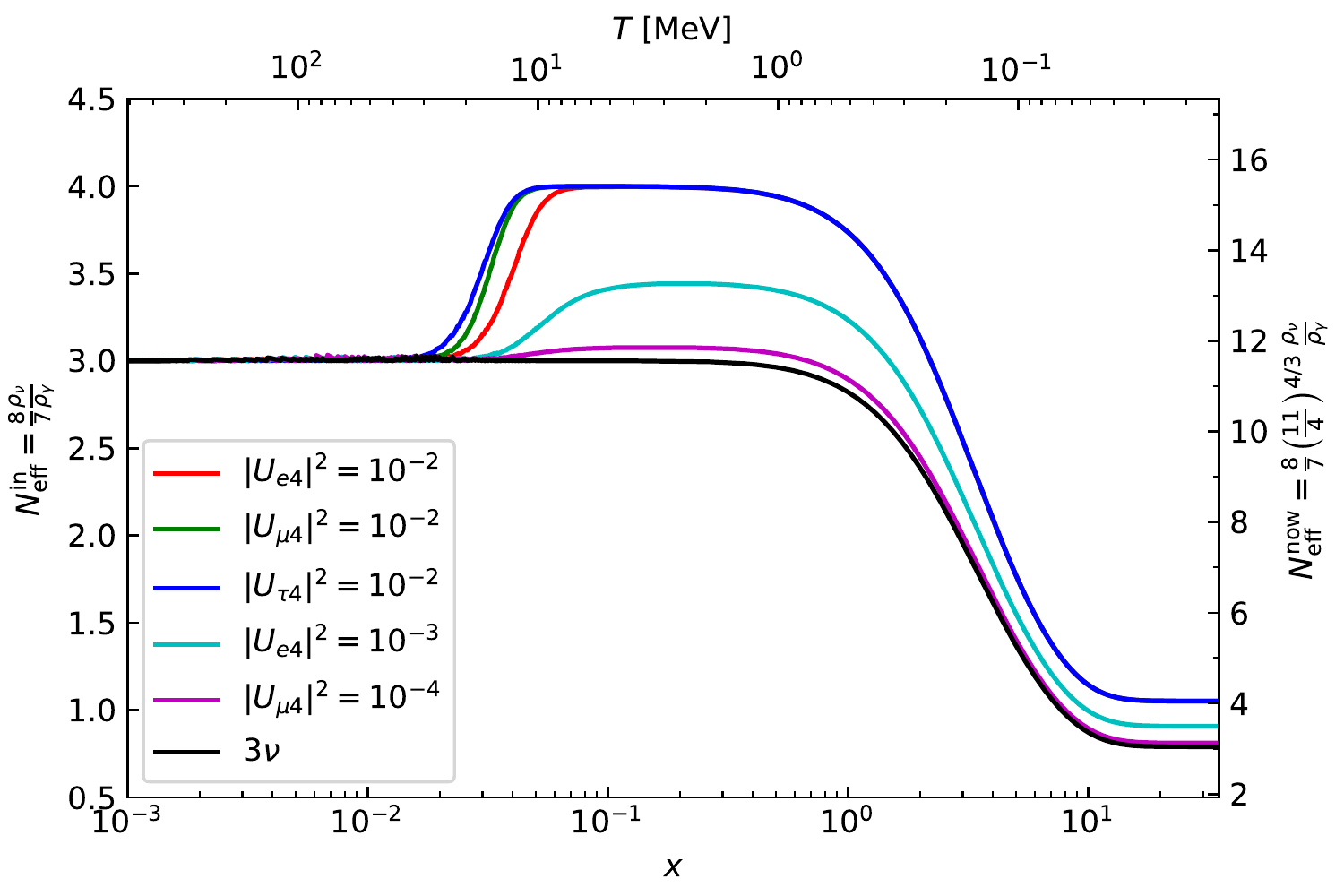}
\caption{\label{fig:Neff_3p1}
Evolution of $\Neff$ in the absence of active-sterile mixing and for some 3+1 cases
(all using $\dmsq{41}=1.29$~eV$^2$ \cite{Gariazzo:2018mwd}, the angles not specified in the legend are set to zero).
One of the 3+1 cases corresponds to $\Neff\simeq3.5$ (cyan),
one to $\Neff\simeq3.13$ (violet) and
the others to $\Neff\simeq4.05$.
Since the number of effective neutrino species is not well defined in the intermediate range,
we report in separate scales the value at early times $\Neff^{\rm in}$ (left axis)
and at late times $\Neff^{\rm now}$ (right axis).
}
\end{figure}

The thermalisation of a new relativistic particle in the early Universe
affects the energy density of radiation. This is usually quantified with the
effective number of neutrino species, \Neff, that is defined before
neutrino decoupling as
\begin{equation}
\Neff^{\rm in}
\equiv
\frac{8}{7}
\frac{\rho_\nu}{\rho_\gamma}\,,
\end{equation}
while after photons have been heated by electron-positron annihilations
we have
\begin{equation}
\Neff^{\rm now}=
\frac{8}{7}
\left(\frac{11}{4}\right)^{4/3}
\frac{\rho_\nu}{\rho_\gamma}\,.
\end{equation}

The evolution of the effective number of neutrinos is reported in Fig.~\ref{fig:Neff_3p1} for a few representative cases.
One can see from this figure that the final \Neff\ is slightly larger than three
when only active neutrinos are considered (black curve),
but can grow up to four when an additional state is present.
For the 3+1 cases in the plot we consider again the benchmark mass splitting $\dmsq{41}=1.29$ eV$^2$,
but consider several choices for the mixing angles.
The red curve in the plot represents the case $\uasq{e}=0.01$,
which also arises as the preferred value from the combination of DANSS and NEOS results,
while the other two mixing matrix elements $\uasq{\mu}$ and $\uasq{\tau}$ are fixed to zero.
In such case, the final value of the effective number of neutrinos is very close to $4$ (actually $\Neff\simeq 4.05$).
A similar final result is obtained 
when we use one of the other two angles instead (blue and green curves), but 
the thermalisation of sterile neutrinos occurs slightly before.
Finally, the last two cases shown only lead to incomplete thermalisation: the mixing is not large 
enough to allow a full energy transfer between the sterile and the active states. Thus, when the active neutrinos decouple
from the rest of the plasma the sterile state is not fully populated. For instance, we get 
$\Neff\simeq3.5$ for $\uasq{e}=0.001$ (cyan line), which is just outside the 3$\sigma$ allowed region by DANSS and NEOS
\cite{Gariazzo:2018mwd,Dentler:2017tkw,Dentler:2018sju}.

\subsection{Momentum distributions of the neutrinos}

\begin{figure}[t]
\centering
\includegraphics[width=0.8\textwidth]{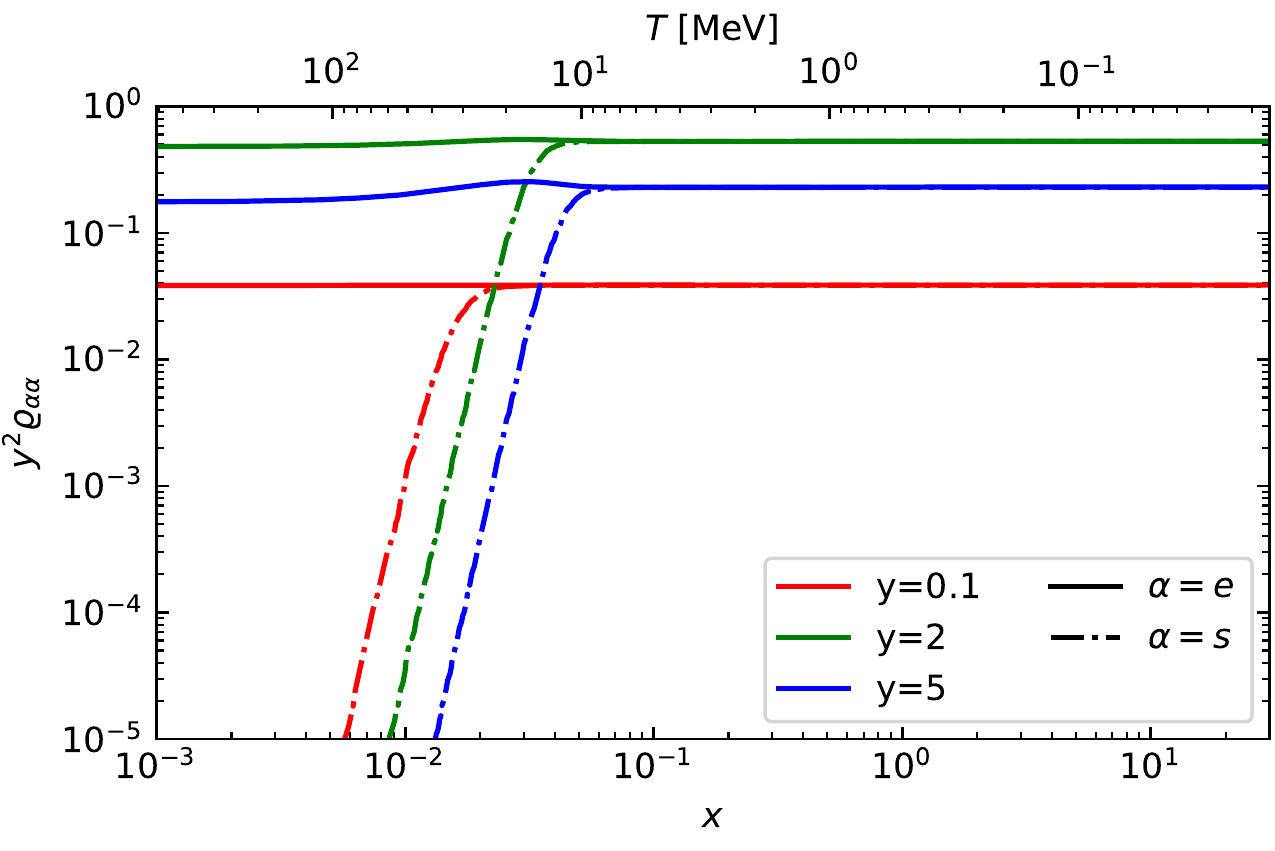}
\caption{\label{fig:rhoii_y_3p1}
Evolution of $\varrho_{\alpha \alpha}$ for different values of $y$ as a function of $x$,
for the $ee$ and $ss$ diagonal entries.
We consider the 3+1 case
using $\dmsq{41}=1.29$~eV$^2$,
$\uasq{e}=0.01$ \cite{Gariazzo:2018mwd}, $\uasq{\mu}=\uasq{\tau}=0$.
}
\end{figure}

The energy distribution function of the sterile neutrino is initially empty, when oscillations are still suppressed.
Once they become effective, the first sterile neutrinos to appear possess very small momenta
and start to reach thermal equilibrium with the active neutrinos. This can be seen in Fig.\ \ref{fig:rhoii_y_3p1},
where we show the evolution of the momentum distribution functions of sterile (dashed-dotted lines) 
and electron (solid lines) neutrinos for different values of the comoving momentum $y$.
As anticipated, we can see that the lowest momenta are populated at earlier times.
It is also interesting to note that the increment in the neutrino temperature due to muon annihilations
is simultaneous to the creation of sterile neutrinos
(visible as a bump in the blue curve at $0.01\lesssim x\lesssim0.1$).

\begin{figure}[t]
\centering
\includegraphics[width=0.8\textwidth]{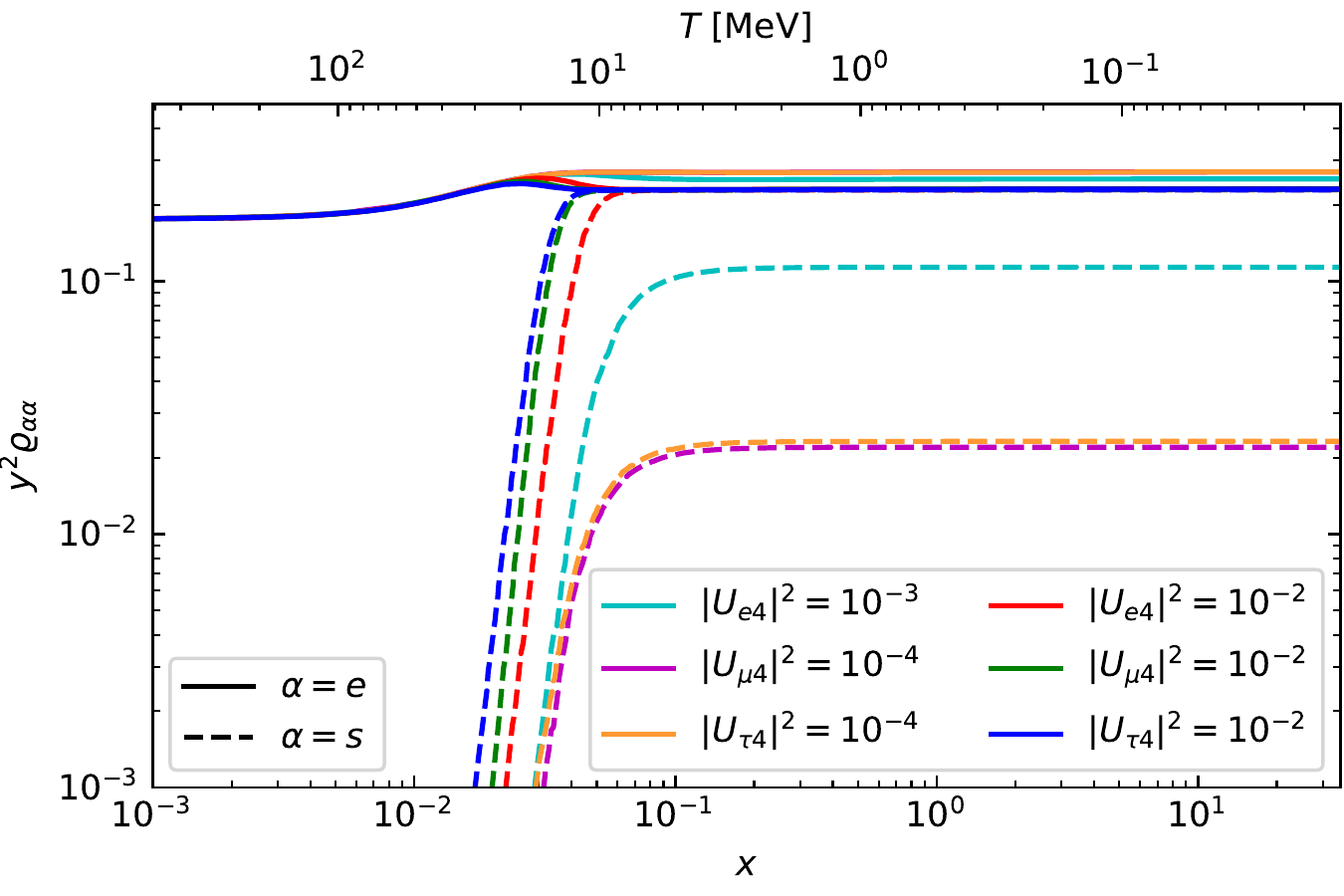}
\caption{\label{fig:rhoii_3p1}
Evolution of $\varrho_{\alpha\alpha}$ for $y=5$ as a function of $x$,
for various cases within the 3+1 scheme. The mass splitting
is always $\dmsq{41}=1.29$~eV$^2$ \cite{Gariazzo:2018mwd},
while the angles not specified in the legend are set to zero.
The 3+1 cases lead to
$\DNeff=1.01$ (red, green and blue),
$\DNeff=0.45$ (cyan) and
$\DNeff\simeq0.10$ (violet and orange).
}
\end{figure}

A similar comparison is shown in Fig.~\ref{fig:rhoii_3p1}, where we now depict the evolution
of the distribution functions at different values of the mixing angles,
all for the same comoving momentum $y=5$.
When the mixing is not large enough, oscillations start later, so that the sterile neutrino does not have time to reach equilibrium
and its momentum distribution remains at a fraction of that of active neutrinos.
For the same final values of \Neff\ (see for instance those that lead to $\Neff\simeq4.05$),
the thermalisation may be slightly different (see $0.01\lesssim x\lesssim0.1$),
but the equilibrium distribution is the same.

\begin{figure}[t]
\centering
\includegraphics[width=0.8\textwidth]{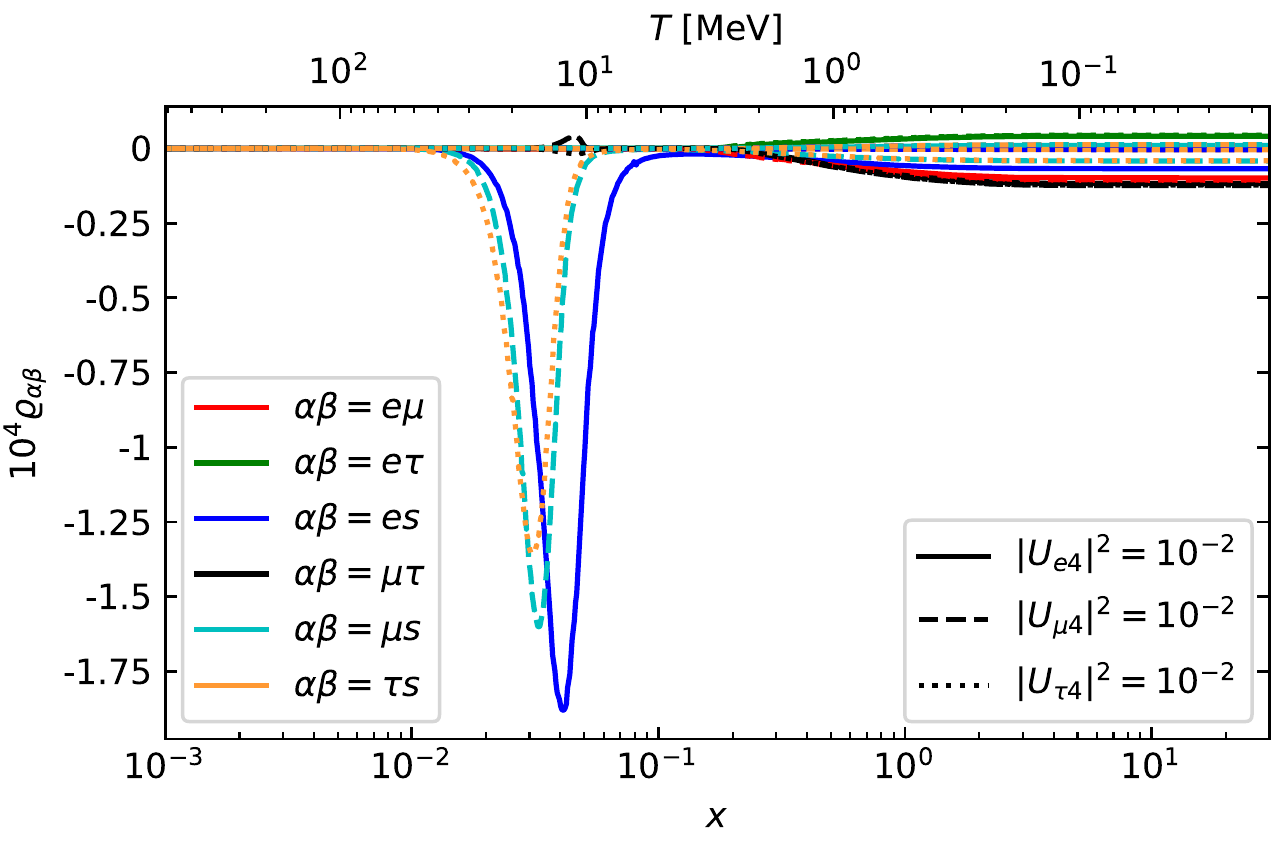}
\caption{\label{fig:rhoij_3p1}
Evolution of $\varrho_{\alpha\beta}$ for $y=5$ as a function of $x$,
for the different off-diagonal entries of various cases with 3+1 neutrinos.
We always have $\dmsq{41}=1.29$~eV$^2$ \cite{Gariazzo:2018mwd},
while the angles not specified in the legend are set to zero.
All the 3+1 cases correspond to $\DNeff=1.01$.
}
\end{figure}

Until now we have discussed only the diagonal elements of the density matrix.
The most interesting things, however, appear in the off-diagonal entries,
which are responsible for the energy transfer between active and sterile neutrinos.
In Fig.~\ref{fig:rhoij_3p1} we show the evolution of the off-diagonal components at $y=5$ for three selected cases,
for which only one among the matrix elements (\uasq{e}, \uasq{\mu} or \uasq{\tau})
is different from zero. It can be easily seen that a strong resonance in the corresponding density matrix entry occurs
just before the diagonal entry is populated.
When \uasq{e} is non zero, for example, the energy is first transferred
from the $\varrho_{ee}$ to the $\varrho_{es}$ component,
and from there to the diagonal entry $\varrho_{ss}$.
At early times, the presence of effective weak interactions makes possible to restore the $\varrho_{es}$ to zero
at the end of the resonance, but this does not happen at late times,
when the various $\varrho_{\alpha\beta}$ remain different from zero,
with an absolute value that dependens on the leading standard mixing angle.

\begin{figure}[t]
\centering
\includegraphics[width=0.8\textwidth]{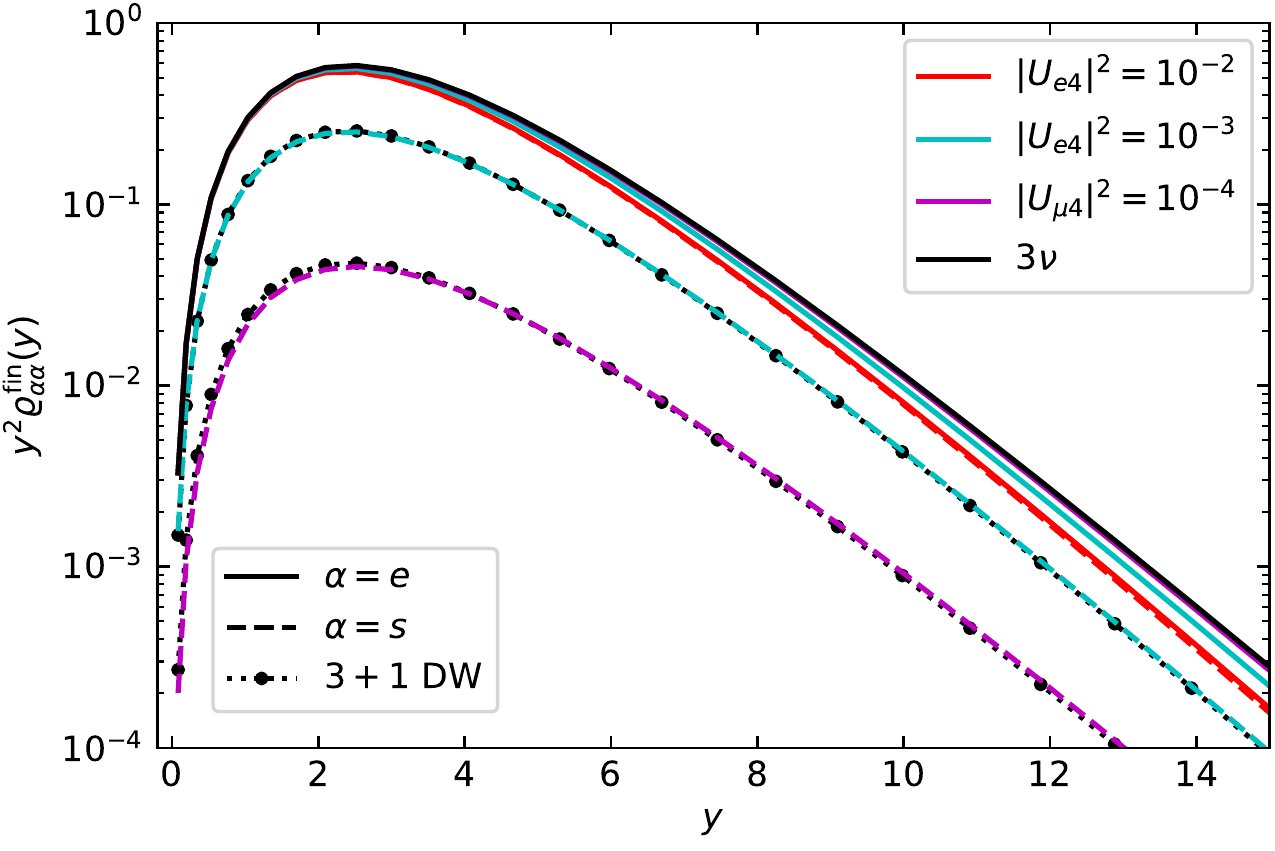}
\caption{\label{fig:rhofinii_3p1}
Final neutrino momentum distribution $\varrho_{\alpha\alpha}$,
for the different diagonal entries.
We show both the standard active neutrino case (black) and some 3+1 cases
(all with $\dmsq{41}=1.29$~eV$^2$ and matrix elements $\uasq{e}=\uasq{\mu}=\uasq{\tau}=0$ unless stated otherwise).
The 3+1 cases correspond to $\DNeff=1.01$ ($\uasq{e}=10^{-2}$, red), $\DNeff=0.45$ ($\uasq{e}=10^{-3}$, cyan)
and $\DNeff\simeq0.10$ ($\uasq{\mu}=10^{-4}$, violet), respectively.
The black dashed line with dots represents the analytical expression in the Dodelson-Widrow (DW) approximation \cite{Dodelson:1993je}
with the neutrino temperature and \DNeff\ obtained from the corresponding coloured curves.
}
\end{figure}

Finally, let us analyse the final shape of the momentum distribution functions after the neutrino decoupling is complete.
In Fig.~\ref{fig:rhofinii_3p1} we show the final diagonal elements of the density matrix, multiplied by a factor $y^2$
(the quantity that is integrated in the calculation of the number density of neutrinos).
Several cases are shown in the figure, corresponding to different assumptions about the neutrino sector.
The black curve is computed using only active neutrinos,
while the red, cyan and violet ones are obtained fixing $\uasq{e}=10^{-2}$, $\uasq{e}=10^{-3}$ and $|U_{\mu 4}|^2 = 10^{-4}$, respectively,
being the other active-sterile angles fixed to zero.
These cases lead to $\Neff\simeq4.05$, $\Neff\simeq3.5$ and $\Neff\simeq3.13$,
so the cyan and violet curves correspond to a sterile neutrino not fully thermalised.
The difference between the black and the red curves, when the sterile is completely thermalised, is only due to the fact that the neutrino temperature
is smaller when part of the energy is transferred to the sterile state.
Instead, the $\uasq{e}=10^{-3}$ or $\uasq{\mu}=10^{-4}$ cases are more interesting, because 
the sterile state shares the same temperature of the active neutrino
and its momentum distribution function is simply rescaled by a factor \DNeff.
To check this, we plot the analytic expression obtained using $f(y)=\DNeff\,[\exp(y/w)+1]^{-1}$
over the points of the $y$ grid that we used. The analytic expression matches reasonably well the computed result.

The above expression for $f$ corresponds to what was found by Dodelson and Widrow \cite{Dodelson:1993je} (DW),
who derived it assuming non-resonant oscillations between two neutrino states,
and is usually adopted as one of the two possible ways to parametrise the momentum distribution function
of the sterile neutrino in phenomenological analyses
(see e.g.~\cite{Acero:2008rh,Gariazzo:2013gua}).
The other possibility is to assign to the sterile neutrino a momentum distribution that has a different
temperature with respect to the one of active neutrinos, i.e.\ $f(y)=[\exp(y/w_s)+1]^{-1}$,
where $w_s\equiv T_s\,a = (T_s/T_\nu)\,w$ and $T_s/T_\nu=\DNeff^{1/4}$
(see e.g.~\cite{Acero:2008rh,Gariazzo:2013gua,Bridle:2016isd,Knee:2018rvj,Berryman:2019nvr}).
Thus, we show that this latter possibility, denoted as {\it thermal distribution}, is 
inaccurate to study the 3+1 sterile neutrino case, for which we find that the DW approximation is excellent.

\subsection{Dependence on the active-sterile mixing parameters}
\label{ssec:neff_mixing}

\begin{figure}[tp]
\centering
\includegraphics[width=0.8\textwidth]{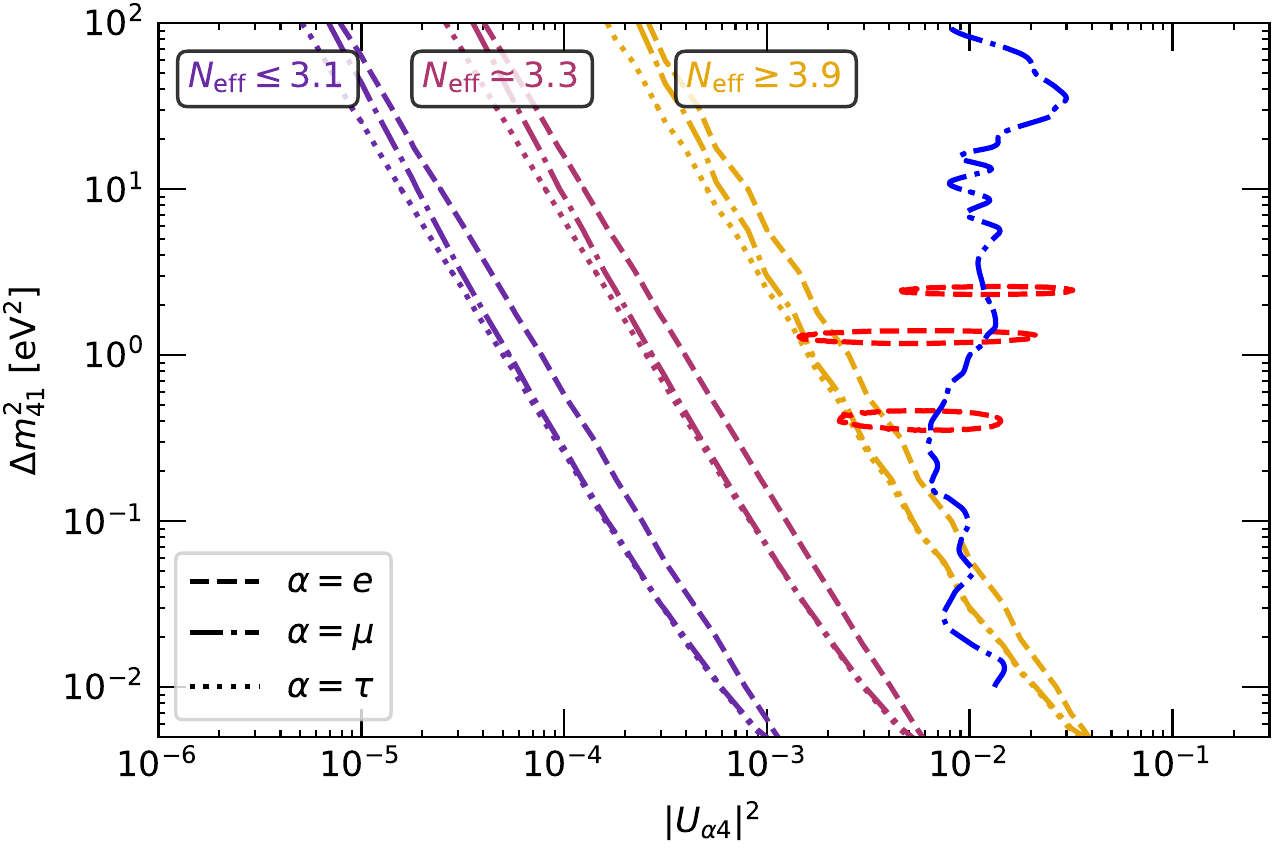}
\includegraphics[width=0.8\textwidth]{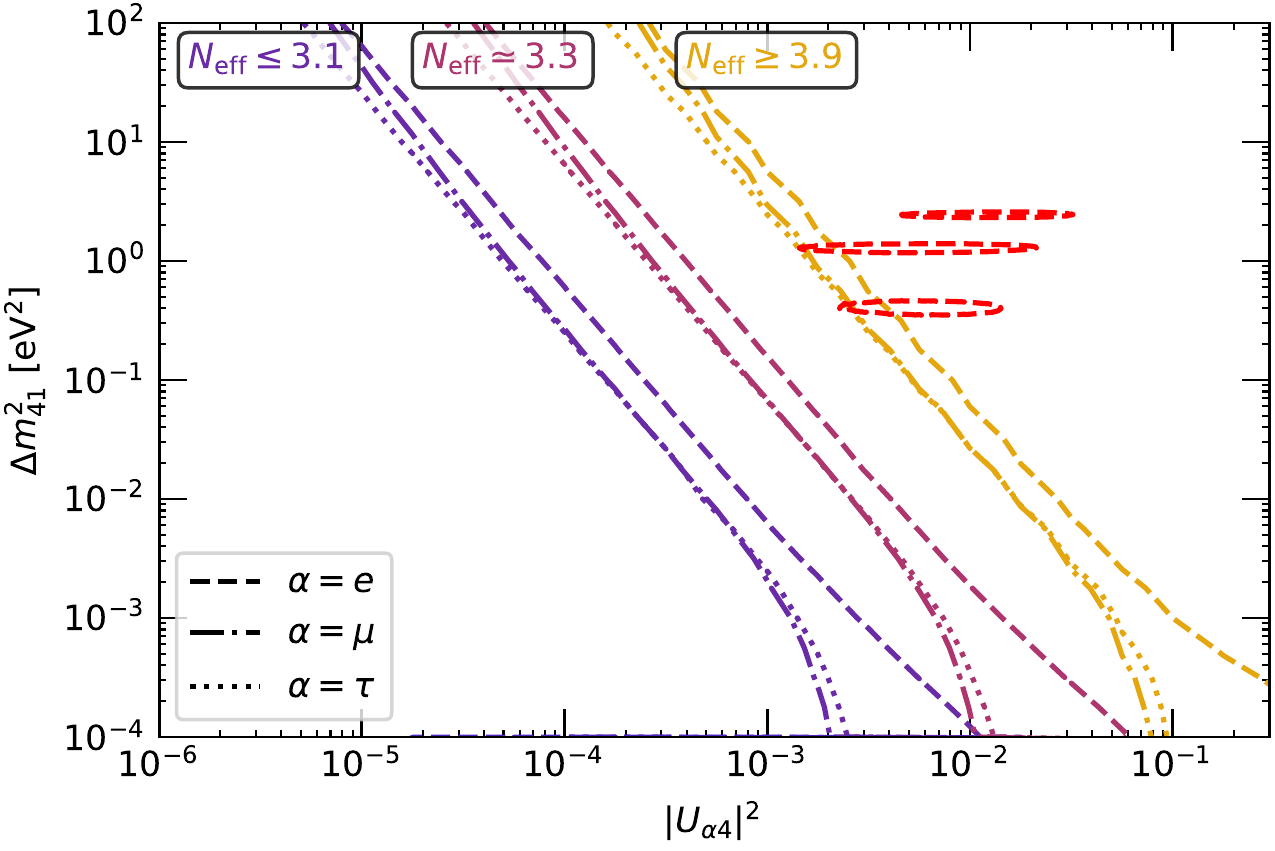}
\caption{\label{fig:Neff_Ui4sq_dm41_comparison}
Comparison of the effect on \Neff\ when varying only one of the active-sterile angles,
for normal (upper panel) or inverted (lower panel) mass ordering.
Dashed, dashed-dotted and dotted lines indicate that only $\theta_{14}$, $\theta_{24}$ or $\theta_{34}$,
respectively, is different from zero.
The different colours encode three discrete levels of \Neff\ as indicated in the text boxes.
Red lines show the allowed regions (99.7\% CL) from DANSS+NEOS \cite{Gariazzo:2018mwd} on $\uasq{e}$,
while blue lines show the constraints (99.7\% CL) from muon (anti)neutrino disappearance
\cite{Gariazzo:inprep} on $\uasq{\mu}$. 
}
\end{figure}

Let us now discuss the main novelty of our paper, namely the different impact of the three active-sterile mixing angles
$\theta_{14}$, $\theta_{24}$ and $\theta_{34}$ on the thermalisation process
when considering the full oscillation paradigm of the 3+1 scheme. We anticipate that a general conclusion is the following:
the effect of the three angles is quite similar but not exactly equal.

A first comparison can be performed in the case when only one of the mixing angles is non zero. 
We show\footnote{Since we consider the sterile neutrino in a 3+1 scheme,
we restrict ourselves to the case $m_4>m_3$ (normal ordering) or $m_4>m_2$ (inverted ordering),
i.e.\ $\dmsq{41}>\dmsq{31}\simeq 2.5\e{-3}\mbox{ eV}^2$ and
$\dmsq{41}>\dmsq{21}\simeq 7.5\e{-5}\mbox{ eV}^2$, respectively. Thus,
the new squared mass difference does not significantly alter the oscillations of the three active neutrinos.} in
Fig.~\ref{fig:Neff_Ui4sq_dm41_comparison} the iso-\Neff\ contours obtained
when varying only $\theta_{14}$, $\theta_{24}$ or $\theta_{34}$. One can see
that, for the same value of the mixing parameter, $\theta_{14}$ always corresponds to the smallest final \Neff.
In other words, a larger $\theta_{14}$ is required to achieve the same \Neff\ with respect to $\theta_{24}$ or $\theta_{34}$.
These latter two angles have a very similar effect at small \dmsq{41},
but $\theta_{34}$ is slightly more effective for larger \dmsq{41}.
This is due to the fact that when $\theta_{24}$ is not zero, the thermalisation is mainly generated by
$\nu_\mu\leftrightarrow\nu_s$ oscillations which, at high temperatures, are
affected by the matter potential created by the few muons that are still present
in the plasma, therefore slowing down the population of the sterile states.

\begin{figure}[tp]
\begin{tabular}{cc}
\multicolumn{2}{c}{\includegraphics[width=0.7\textwidth]{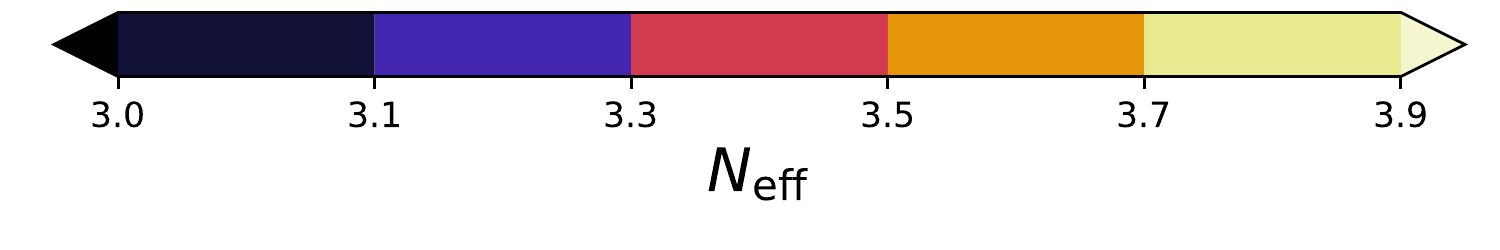}}\\
\includegraphics[width=0.48\textwidth]{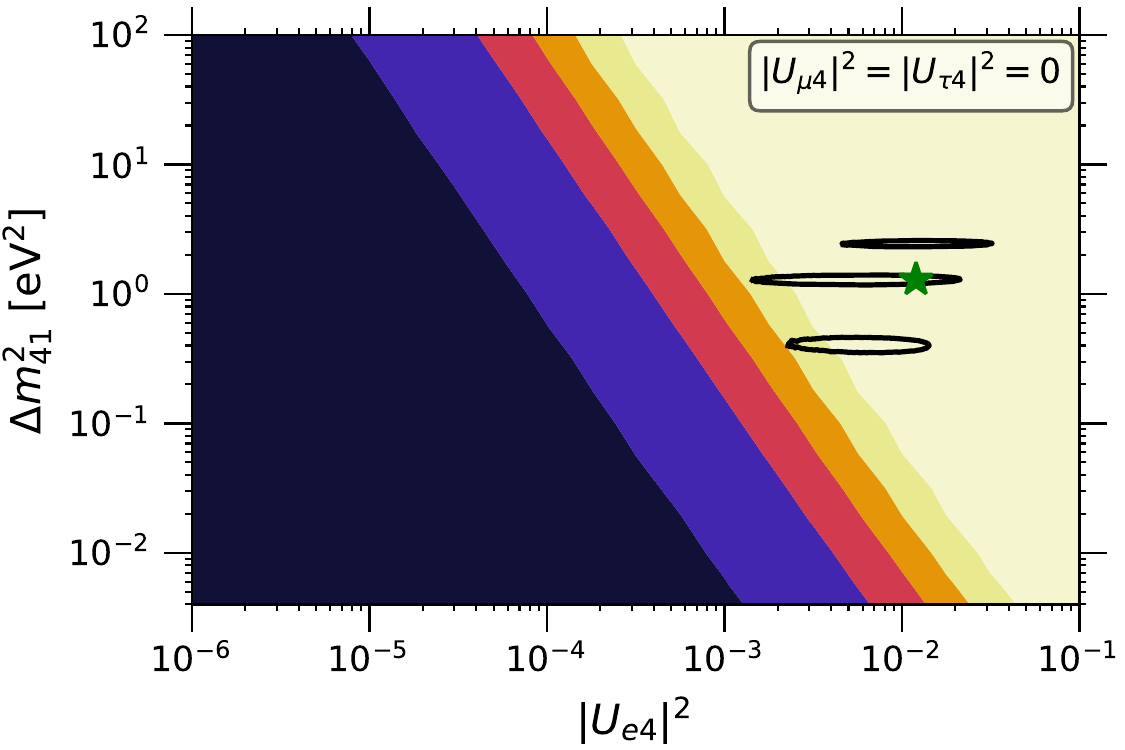}&
\includegraphics[width=0.48\textwidth]{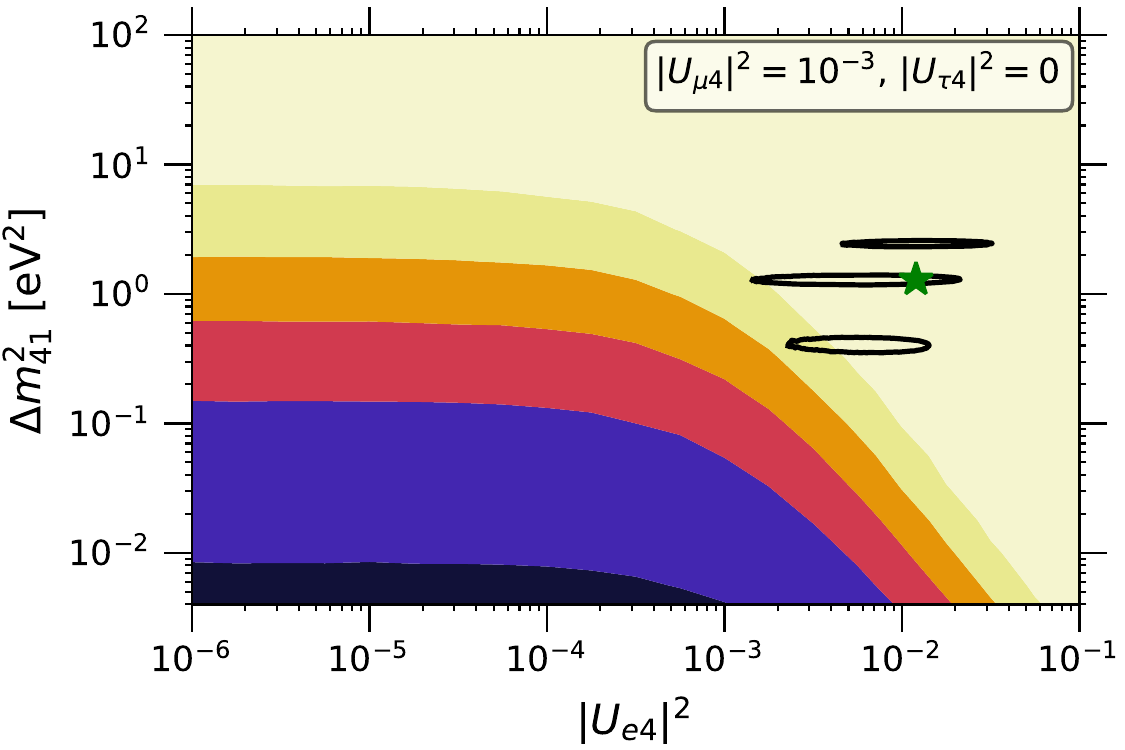}\\
\includegraphics[width=0.48\textwidth]{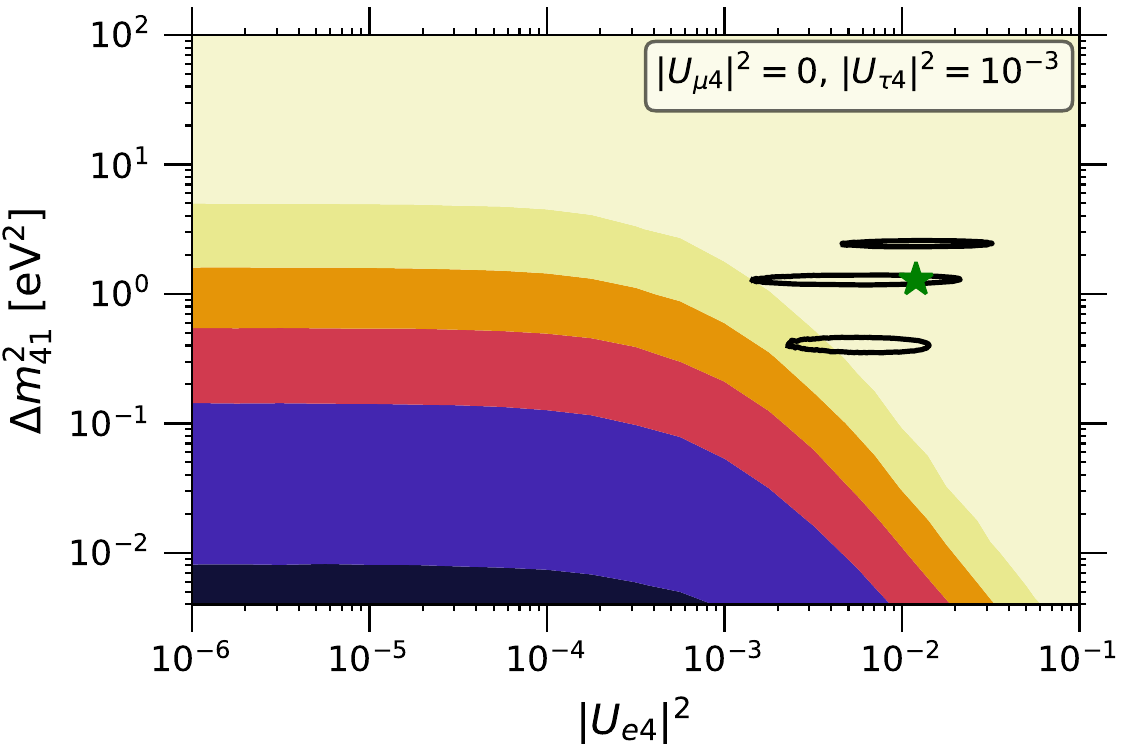}&
\includegraphics[width=0.48\textwidth]{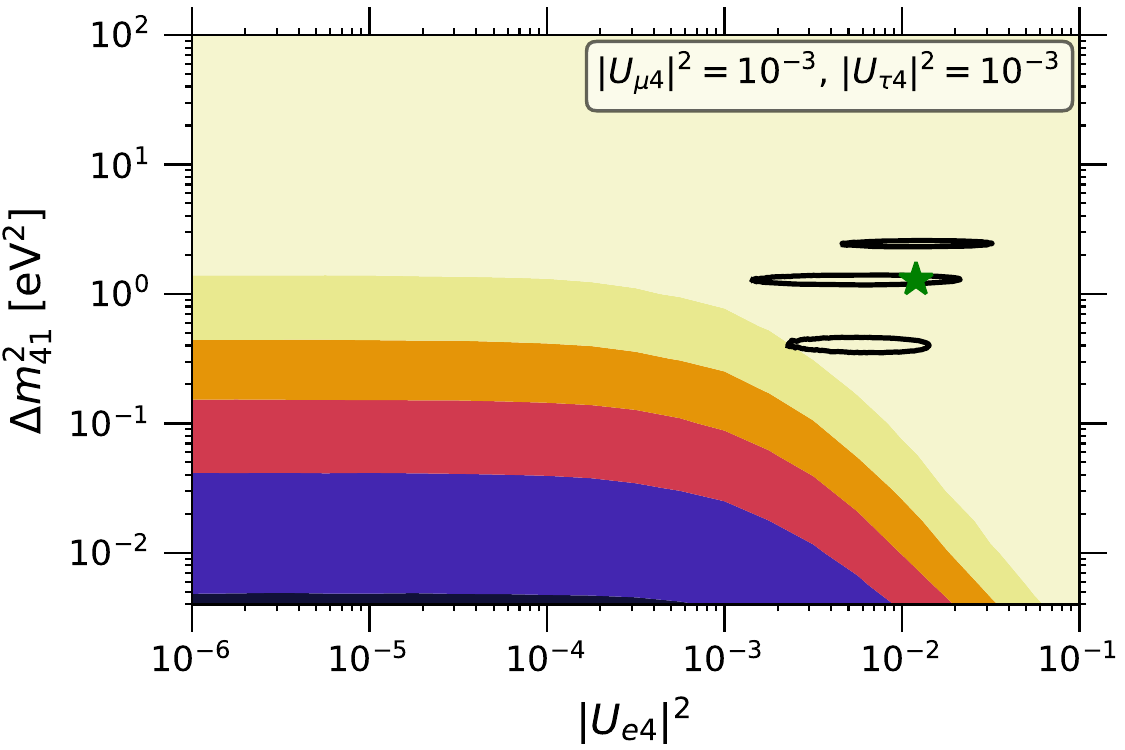}
\end{tabular}
\caption{\label{fig:Neff_Ui4sq_dm41}
Final \Neff\ in the 3+1 case for different values of \dmsq{41} 
and \uasq{e} when considering normal ordering for the active neutrinos. The
other two active-sterile components of the mixing matrix take the values as labelled.
The black closed contours represent the 3$\sigma$ preferred regions
and the green star the best-fit point from \cite{Gariazzo:2018mwd}.
}
\end{figure}

In the same figure~\ref{fig:Neff_Ui4sq_dm41_comparison} we can also compare the $\Neff\geq3.9$ lines
with the preferred regions for $\uasq{e}$ at 99.7\% CL
from DANSS+NEOS \cite{Gariazzo:2018mwd,Dentler:2018sju,Dentler:2017tkw} 
and the exclusion curves for $\uasq{\mu}$ at 99.7\% CL from muon (anti)neutrino disappearance \cite{Gariazzo:inprep}.
Note that the blue curve is not shown in the bottom panel, corresponding to inverted mass ordering for the active neutrinos,
as it is derived under the assumption of normal ordering.
While the DANSS+NEOS region is also derived assuming normal ordering, it does not extend to regions
where \dmsq{41} becomes comparable to $|\dmsq{31}|$,
and is therefore valid in both cases. The iso-\Neff\ contours in Fig.~\ref{fig:Neff_Ui4sq_dm41_comparison}
can be very well approximated by straight lines for each mixing angle, as shown in
previous analyses, see e.g.~\cite{Dolgov:2003sg,Hannestad:2012ky,Bridle:2016isd}. 
In particular, our results for the 3+1 case are in reasonable agreement (within few percent of the total \Neff) with those obtained 
with the \texttt{LASAGNA} code in the 1+1 approximation.

In the following,
let us consider what happens when we increase the values of the angles that were earlier always fixed to zero.
An example is shown in the four panels of Fig.~\ref{fig:Neff_Ui4sq_dm41}. 
The iso-\Neff\ contours change when we vary \dmsq{41} and \uasq{e}
while the two remaining matrix elements \uasq{\mu} or \uasq{\tau} assume different values.
It is interesting that these contours remain similar to those in Fig.~\ref{fig:Neff_Ui4sq_dm41_comparison}
when the largest mixing comes from \uasq{e}, but saturate as a consequence of the other mixing channels
when \uasq{e} is smaller than one of the other two mixing matrix elements. 
We include in the same panels
the preferred 99.7\% CL regions by DANSS+NEOS \cite{Gariazzo:2018mwd}.
One can conclude that the current preferred value for \uasq{e} would lead to a contribution of $\Neff\simeq4$,
regardless of the values of \uasq{\mu} or \uasq{\tau} and despite the fact that $\theta_{14}$
is the angle which makes the thermalisation less effective.
In light of current cos\-mo\-lo\-gical constraints, which prefer 
$\Neff\lesssim3.3$~\cite{Aghanim:2018eyx} (Planck data TT, TE, EE+lowE+lensing+BAO, 95\% CL),
this indicates a strong tension between CMB observations and neutrino oscillation experiments,
as noted in many previous analyses.

\begin{figure}[htp]
\centering
\includegraphics[width=0.63\textwidth]{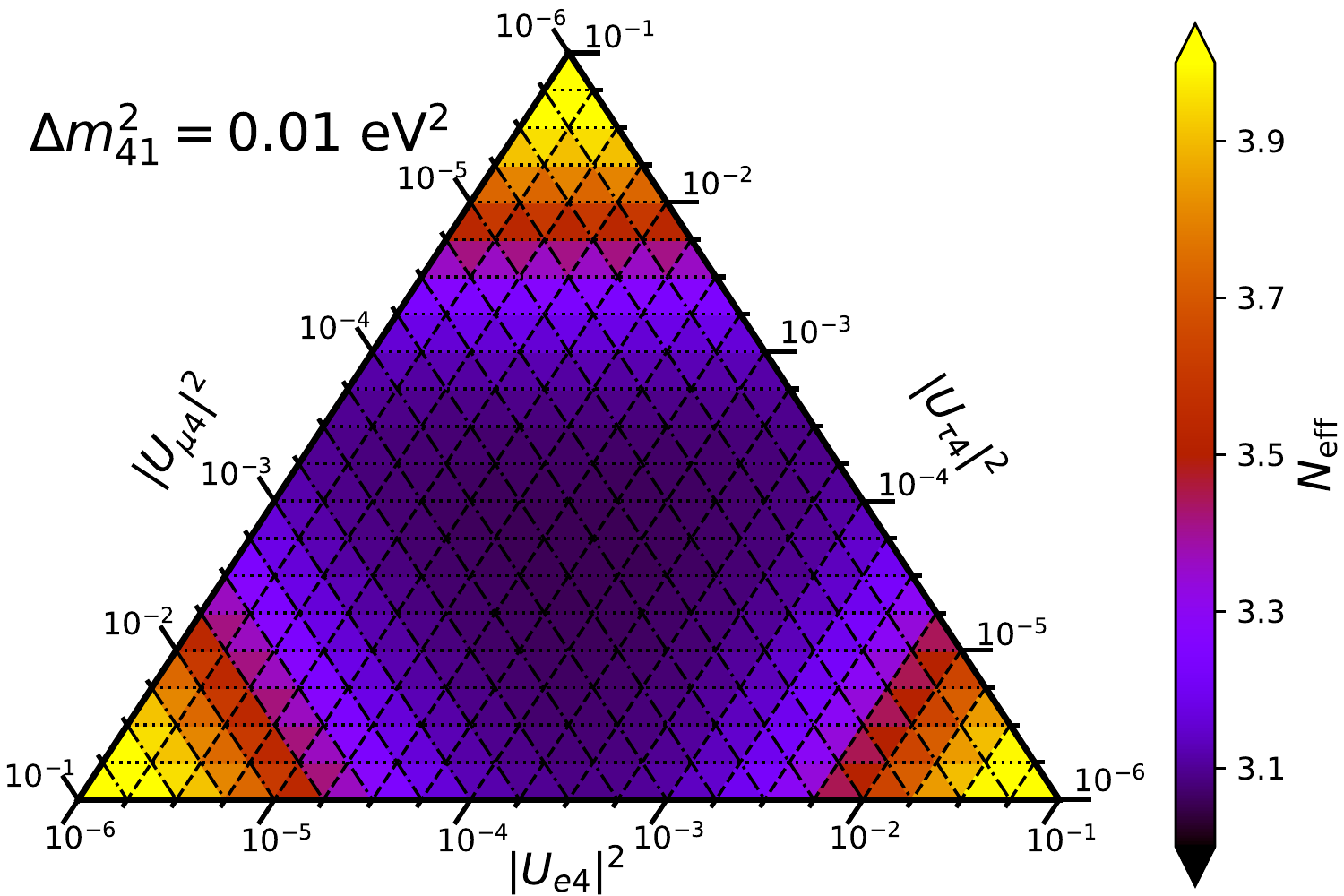}
\includegraphics[width=0.63\textwidth]{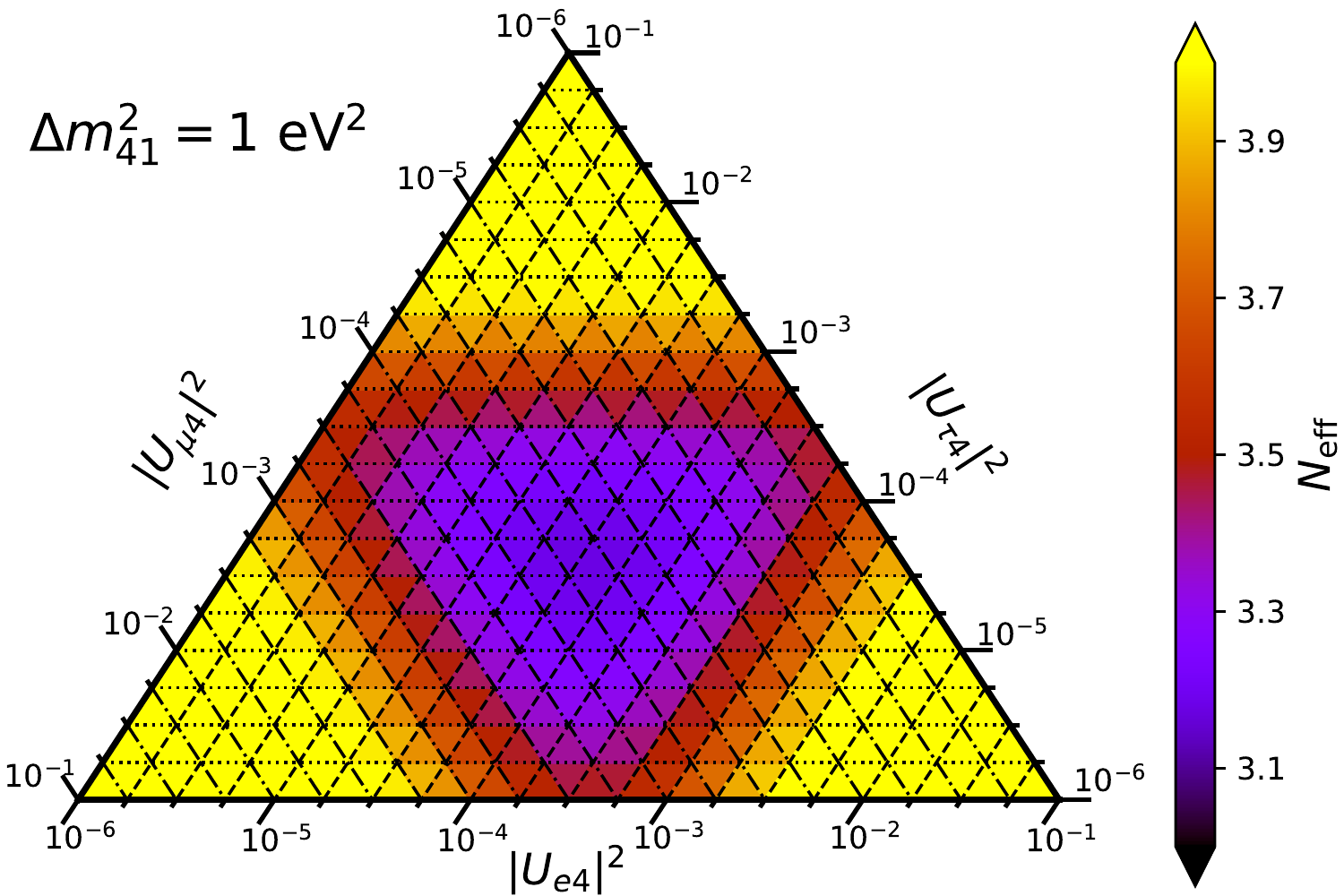}
\includegraphics[width=0.63\textwidth]{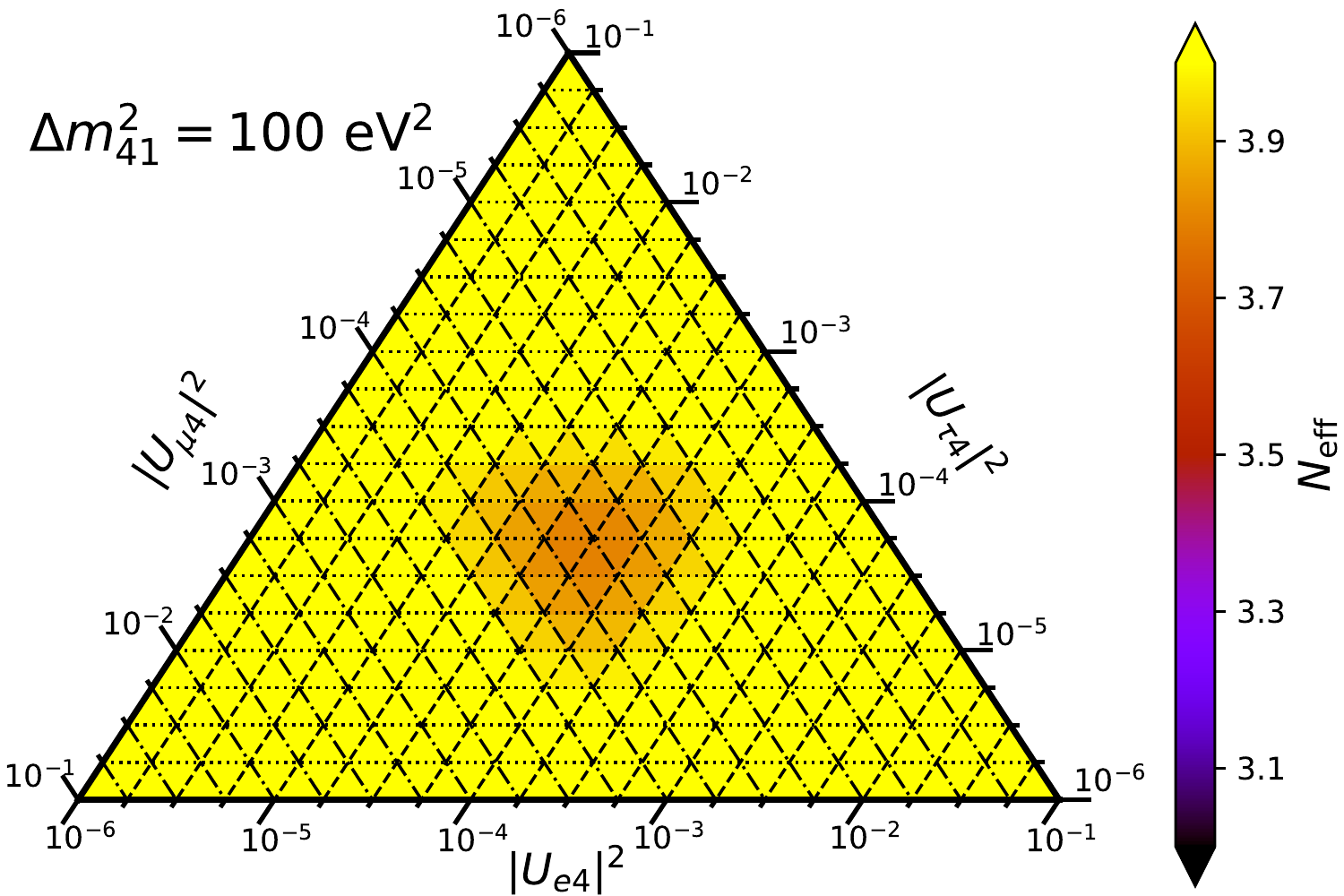}
\caption{\label{fig:ternary_example}
Final \Neff\ in the 3+1 case for different values of \dmsq{41} 
if all three active-sterile mixing angles are varied simultaneously (under the constraint
$\sum\log_{10}|U_{i4}|^2=-13$). Plot produced using a modified version of the routines from \texttt{python-ternary}~\cite{python_ternary}.
}
\end{figure}

Finally, let us discuss in more detail the simultaneous effect of all three active-sterile mixing angles.
To do so, we propose an adapted version of the ternary plot which is sometimes shown for discussing 
the flavour composition of high-energy neutrino fluxes, see e.g.~\cite{Aartsen:2015knd}.
In Fig.~\ref{fig:ternary_example} we show, for three selected values of \dmsq{41},
the effect on the final value of \Neff\ when all three mixing angles are non zero.
Instead of a proper ternary plot,
for which we should have fixed $\uasq{e}+\uasq{\mu}+\uasq{\tau}=1$, we show combinations of the mixing matrix elements such that
$\sum\log_{10}|U_{i4}|^2=-13$. With this choice we avoid the use of a linear scale in the mixing matrix elements that 
would make the plot mostly filled with only $\Neff\simeq4$.
The panels in Fig.~\ref{fig:ternary_example} reflect the fact that the thermalisation is more effective 
when the mass splitting grows, because oscillations start earlier and have more time to develop.
Although these plots look really symmetric, a more accurate inspection of the upper panel shows that the centre of the darker region (corresponding to small \Neff)
is not located exactly at the centre of the triangle, a consequence of the different interactions and masses of the three active neutrinos.

\section{Conclusions}
\label{sec:conc}

The existence of a fourth neutrino state, insensitive to weak interactions but mixed with the three active neutrinos, could provide an explanation
for the anomalies found in some short-baseline oscillation experiments. While we expect that new oscillation data will confirm or refute this solution, it is interesting to obtain complementary information from other terrestrial, astrophysical and cosmological observations. In this regard, it is important to calculate the contribution of sterile neutrinos to the cosmological energy density of radiation, since it is well constrained by present data, for instance from CMB measurements. 

We have presented in this paper a novel calculation of the thermalisation of the mostly sterile neutrino state in the early Universe
in the 3+1 scheme, solving the momentum-dependent kinetic equations for the distribution functions. 
For the first time, we consider the simultaneous effect of all mixing angles involved in neutrino oscillations, fixing those exclusive of 
active neutrinos to the values obtained in global-fit analyses but leaving the three active-sterile angles as free parameters.
For this purpose we have developed a new numerical code, \fortepiano\footnote{This code will be publicly available at \url{https://bitbucket.org/ahep_cosmo/fortepiano_public}.}, that follows the evolution of the Boltzmann equations 
in a momentum-grid basis, accounting for the full oscillation mixing matrix, the relevant processes of weak interactions 
and the expansion of the Universe. 

We have studied the evolution of the comoving energy and entropy densities of the elementary particles in 
the cosmic plasma, from the epoch of muon-antimuon pair annihilations until the decoupling process of active neutrinos is complete.
We show that the population of sterile states when flavour oscillations become effective slightly increases the entropy density of the relativistic plasma 
and reduces its comoving temperature with respect to the standard three-neutrino case. The final energy density of the plasma is 
also slightly reduced in the 3+1 scenario. Our code also provides the evolution of the momentum distributions of each neutrino state.
In particular, we show that the so-called Dodelson-Widrow approximation \cite{Dodelson:1993je} for the final energy spectrum of 
sterile neutrinos is in very good agreement with our full calculations, while the thermal distribution approximation is inaccurate.

Our main results concern the dependence of the final value of $N_{\rm eff}$ on the mixing pa\-ra\-meters of the 3+1 scheme
(the three new mixing angles and the squared mass difference $\Delta m_{41}^2$). Focusing on the effect of one single angle,
i.e.\ neglecting the other two, for the same $\Delta m_{41}^2$  the angle $\theta_{14}$ leads always to smaller $N_{\rm eff}$
due to the delay in the thermalisation process of the mostly sterile state caused by the matter potential from electrons and positrons 
in the plasma. Instead, the individual effect of $\theta_{24}$ or $\theta_{34}$ is similar but not equal. We have also shown how
the value of $N_{\rm eff}$ is modified when all three mixing angles possess non-zero values. While we get $N_{\rm eff} = 3.044$ in the absence of 
active-sterile mixing, in agreement with previous calculations, we find, as expected, that the $3\sigma$ preferred region from the 
analysis of oscillation data from NEOS + DANSS leads to a value of $N_{\rm eff}\simeq 4$, in tension with the cosmological bounds.
Of course, when more than one active-sterile mixing angle is included the tension is enlarged,
since the presence of more channels favours the population of sterile neutrinos. Thus, if the existence of active-sterile mixing is confirmed,
a new ingredient to suppress the thermalisation of the fourth neutrino state will be required, such as the presence of a 
neutrino-antineutrino asymmetry (see e.g.\ \cite{Hannestad:2012ky,Saviano:2013ktj})
or secret neutrino interactions \cite{Archidiacono:2016kkh,Forastieri:2017oma}.

In conclusion, our analysis is a new step towards understanding better the effects of active-sterile oscillations in the early Universe.
While the obtained values of \Neff\ can be directly compared with the preferred range of this parameter from cosmological fits 
(except for the larger values of $\Delta m_{41}^2$, for which neutrinos become non-relativistic at the relevant epochs), in the future we plan
to extend this work performing a detailed calculation of the bounds on the active-sterile mixing parameters from the full
set of oscillation data and cosmological measurements, along the lines of \cite{Mirizzi:2013gnd,Bridle:2016isd,Knee:2018rvj,Berryman:2019nvr}.

\acknowledgments
We thank Julien Lesgourgues for suggesting to consider the Gauss-Laguerre quadrature method,
Steffen Hagstotz for discussions on the \texttt{LASAGNA} code
and Carlo Giunti for useful comments. 
Work supported by the Spanish grants SEV-2014-0398 and FPA2017-85216-P (AEI/FEDER, UE), 
PROMETEO/2018/165 (Generalitat Valenciana) and the Red Consolider
MultiDark FPA2017-90566-REDC. SG receives support from the
European Union's Horizon 2020 research and innovation
program under the Marie Sk\l{}odowska-Curie individual
Grant Agreement No.\ 796941.
PFdS acknowledges support by the Vetenskapsr{\aa}det (Swedish Research Council) 
through contract No.\ 638-2013-8993 and the Oskar Klein Centre for Cosmoparticle Physics.
SG and SP thank the Institute for Theoretical Particle Physics and Cosmology of 
RWTH Aachen University for hospitality and support during the final phase of this work.

\appendix

\section{Collision integrals}
\label{sec:collint}
Neutrino interactions are encoded in the matrices of couplings $G^L$ and $G^R$, for left- or right-handed particles,
\begin{equation}
G^L=\text{diag}(g_L, \tilde g_L, \tilde g_L, 0)\,,
\qquad
G^R=\text{diag}(g_R, g_R, g_R, 0)\,,
\end{equation}
where $g_L=\sin^2\theta_W+1/2$, $\tilde g_L=\sin^2\theta_W - 1/2$, $g_R=\sin^2\theta_W$,
and $\theta_W$ is the weak mixing angle.

The full collision terms are defined by the sum
of the contributions from neutrino--electron/positron scattering and
$e^\pm$ annihilation into neutrinos.
We neglect other reactions, such as $\mu^\pm$ annihilation (which only affects at very early temperatures when everything is in equilibrium)
and neutrino--neutrino scattering.
We therefore have \cite{deSalas:2016ztq}
\begin{eqnarray}
\mathcal{I}[\varrho(y)]
&=&
\frac{G_F^2}{(2\pi)^3y^2}
\left(\mathcal{I}_{\rm sc}^u + \mathcal{I}_{\rm ann}^u\right)\,,
\label{eq:collint}
\\
\mathcal{I}_{\rm sc} ^u
&=&
\int {\rm d}y_2 {\rm d}y_3 \frac{y_2}{E_2} \frac{y_4}{E_4}
\label{eq:I_sc}
\\
&&
\left\{\left(\Pi_2^s(y, y_4)+\Pi_2^s(y, y_2)\right)
    \left[
    F_{\rm sc}^{LL}\left(\varrho^{(1)}, f_e^{(2)}, \varrho^{(3)}, f_e^{(4)}\right)
    +F_{\rm sc}^{RR}\left(\varrho^{(1)}, f_e^{(2)}, \varrho^{(3)}, f_e^{(4)}\right)\right]\right.
    \nonumber\\
    &&\left.-2(x^2+\delta m_e^2)\Pi_1^s(y,y_3)
    \left[
     F_{\rm sc}^{RL}\left(\varrho^{(1)}, f_e^{(2)}, \varrho^{(3)}, f_e^{(4)}\right)
    +F_{\rm sc}^{LR}\left(\varrho^{(1)}, f_e^{(2)}, \varrho^{(3)}, f_e^{(4)}\right)
    \right]\nonumber
\right\}\,,
\\
\mathcal{I}_{\rm ann}^u
&=&
\int {\rm d}y_2 {\rm d}y_4 \frac{y_3}{E_3} \frac{y_4}{E_4}
\label{eq:I_ann}\\
    &&\left\{\Pi_2^a(y, y_4)F_{\rm ann}^{LL}\left(\varrho^{(1)}, \varrho^{(2)}, f_e^{(3)}, f_e^{(4)}\right)
    +\Pi_2^a(y, y_3)F_{\rm ann}^{RR}\left(\varrho^{(1)}, \varrho^{(2)}, f_e^{(3)}, f_e^{(4)}\right)\right.
\nonumber\\
    &&\left.+ (x^2+\delta m_e^2)\Pi_1^a(y,y_2)
    \left[
    F_{\rm ann}^{RL}\left(\varrho^{(1)}, \varrho^{(2)}, f_e^{(3)}, f_e^{(4)}\right)
    +F_{\rm ann}^{LR}\left(\varrho^{(1)}, \varrho^{(2)}, f_e^{(3)}, f_{e^+}^{(4)}\right)
    \right]\nonumber
\right\}\,,
\end{eqnarray}
where $E^2_i = \sqrt{x^2+y_i^2+\delta m_e^2}$ and
\begin{eqnarray}
\Pi_1^s(y,y_3)
&=&
y\,y_3\,D_1+D_2(y,y_3,y_2,y_4),
\\
\Pi_1^a(y,y_2)
&=&
y\,y_2\,D_1-D_2(y,y_2,y_3,y_4),
\\
\Pi_2^s(y,y_2)/2
&=&
y\,E_2\,y_3\,E_4\,D_1 + D_3 - y\,E_2 D_2(y_3,y_4,y,y_2) - y_3\,E_4 D_2(y,y_2,y_3,y_4),
\\
\Pi_2^s(y,y_4)/2
&=&
y\,E_2\,y_3\,E_4\,D_1 + D_3 + E_2\,y_3 D_2(y,y_4,y_2,y_3) + y\,E_4 D_2(y_2,y_3,y,y_4),
\\
\Pi_2^a(y,y_3)/2
&=&
y\,y_2\,E_3\,E_4\,D_1 + D_3 + y\,E_3 D_2(y_2,y_4,y,y_3) + y_2\,E_4 D_2(y,y_3,y_2,y_4),
\\
\Pi_2^a(y,y_4)/2
&=&
y\,y_2\,E_3\,E_4\,D_1 + D_3 + y_2\,E_3 D_2(y,y_4,y_2,y_3) + y\,E_4 D_2(y_2,y_3,y,y_4),
\end{eqnarray}
where the functions $D_i$ have the following definitions \cite{Dolgov:1997mb}:
\begin{eqnarray}
D_1(a,b,c,d)
&=&
\frac{16}{\pi}
\int_0^\infty
\frac{{\rm d}\lambda}{\lambda^2}
\prod_{i=a,b,c,d}\sin(\lambda i)
\,,\\
D_2(a,b,c,d)
&=&
-\frac{16}{\pi}
\int_0^\infty
\frac{{\rm d}\lambda}{\lambda^4}
\prod_{i=a,b}\left[\lambda i \cos(\lambda i)-\sin(\lambda i)\right]
\prod_{j=c,d}\sin(\lambda j)
\,,\\
D_3(a,b,c,d)
&=&
\frac{16}{\pi}
\int_0^\infty
\frac{\mathrm{d}\lambda}{\lambda^6}
\prod_{i=a,b,c,d}\left[\lambda i \cos(\lambda i)-\sin(\lambda i)\right]
\,.
\end{eqnarray}
The three functions can be written in a more efficient way for the calculation,
since they can be solved analytically, see e.g.~\cite{Blaschke:2016xxt} for the complete expressions.

Finally, the functions that define the phase space factors in the collision terms are \cite{deSalas:2016ztq}:
\begin{eqnarray}
F_{\rm sc}^{ab}\left(\varrho^{(1)}, f_e^{(2)}, \varrho^{(3)}, f_e^{(4)}\right)
&=&
f_e^{(4)}(1-f_e^{(2)})\left(G^a\varrho^{(3)}G^b(1-\varrho^{(1)})+(1-\varrho^{(1)})G^b\varrho^{(3)}G^a\right)
\nonumber\\
&-&
f_e^{(2)}(1-f_e^{(4)})\left(\varrho^{(1)}G^b(1-\varrho^{(3)})G^a+G^a(1-\varrho^{(3)})G^b\varrho^{(1)}\right),
\label{eq:F_ab_sc}\\
F_{\rm ann}^{ab}\left(\varrho^{(1)}, \varrho^{(2)}, f_e^{(3)}, f_e^{(4)}\right)
&=&
f_e^{(3)}f_e^{(4)}\left(G^a(1-\varrho^{(2)})G^b(1-\varrho^{(1)})+(1-\varrho^{(1)})G^b(1-\varrho^{(2)})G^a\right)
\nonumber\\
&-&
(1-f_e^{(3)})(1-f_e^{(4)})\left(G^a\varrho^{(2)}G^b\varrho^{(1)}+\varrho^{(1)}G^b\varrho^{(2)}G^a\right),
\label{eq:F_ab_ann}
\end{eqnarray}
where $\varrho^{(i)}=\varrho(y_i)$ and $f_e^{(i)}=f_{\rm FD}(y_i, z)$ represent the momentum distribution function
of the various particles.
The full expression for these functions should take into account the lepton asymmetry and distinguish
the momentum distributions of leptons/neutrinos from those of antilepton/antineutrinos.
Since we do not include lepton asymmetry, we just report the expressions without the heavier notation
required to distinguish the various terms.

The code we use can compute the collision terms according to Eqs.~\eqref{eq:I_sc} and \eqref{eq:I_ann},
but the integrals are very expensive.
For the non-diagonal terms of the collision matrix we therefore use the
damping approximation, in the form
\begin{equation}
\mathcal{I}_{\alpha\beta}(\varrho) = -D_{\alpha\beta} \varrho_{\alpha\beta},
\end{equation}
for $\alpha \neq \beta$.
The expressions for the coefficients $D_{\alpha\beta}$ depend on the elements considered.
The coefficients were derived for example in \cite{McKellar:1992ja},
see also \cite{Enqvist:1991qj,Bell:1998ds},
and can be written as
\begin{eqnarray}
D_{e\mu}/F=D_{e\tau}/F & = & 15 + 8\sin^4\theta_W\,,\\
D_{\mu\tau}/F & = & 7 - 4\sin^2\theta_W + 8\sin^4\theta_W\,,\\
D_{es}/F=D_{e\tau}/F & = & 29 + 12\sin^2\theta_W + 24\sin^4\theta_W\,,\\
D_{\mu s}/F = D_{\tau s}/F & = & 29 - 12\sin^2\theta_W + 24\sin^4\theta_W\,,
\end{eqnarray}
where
$F=7\pi^4 y^3/135$ is a common normalisation coefficient.

\section{\fortepiano}
\label{sec:fortepiano}
We present here the main features of
our code, \texttt{FORTran-Evolved PrimordIAl Neutrino Oscillations}
(\fortepiano).
The code will be publicly available at the url
\url{https://bitbucket.org/ahep_cosmo/fortepiano_public}.

\subsection{Equations}
\fortepiano\ was used in this article to compute oscillations with four neutrinos,
but the code is actually written in order to accept up to six neutrinos.
Neutrinos, including the sterile, are always treated as ultra-relativistic particles, which is a good approximation if the
neutrino masses do not exceed $\mathcal{O}(\mbox{a few keV})$,
i.e.\ neutrinos are still fully relativistic at decoupling.
For larger masses, neutrinos may start to become non-relativistic before decoupling,
and in that case one should take into account the effect of the mass.

When using $N$ neutrinos,
the mixing matrix is defined as
\begin{equation}\label{eq:mixing_matrix_nxn}
U=R^{(N-1)N} \ldots R^{1N}
R^{(N-2)(N-1)}\ldots R^{1(N-1)}
\ldots
R^{34} R^{24} R^{14} R^{23} R^{13} R^{12},
\end{equation}
following and extending the convention presented in Eq.~(12) of \cite{Gariazzo:2015rra},
where the rotation matrices are defined as in Eq.~\eqref{eq:rotationmatrix}.
It enters the calculation of the rotated mass matrix
$\mathbb{M}_{\rm F}=U\mathbb{M}U^\dagger$,
where the diagonal mass matrix is
$\mathbb{M}=\text{diag}(m_1^2,\ldots,m_N^2)$.
Other matrices that we need to define are
\begin{equation}\label{eq:matterpotentials_nxn}
\mathbb{E}_\ell=\text{diag}(\rho_e, \rho_\mu, 0, \ldots)\,,
\qquad
\mathbb{E}_\nu=S_a\left(\int \mathrm{d}y y^3\varrho\right) S_a\,
\quad\mbox{with }S_a=\text{diag}(1,1,1,0,\ldots)\,,
\end{equation}
while the interaction matrices become
\begin{equation}
G^L=\text{diag}(g_L, \tilde g_L, \tilde g_L, 0,\ldots)\,,
\qquad
G^R=\text{diag}(g_R, g_R, g_R, 0,\ldots)\,.
\end{equation}
Finally, concerning the collision terms,
we use all the definitions presented in section~\ref{sec:collint},
but with $N\times N$ matrices.
It is easy to see from the definitions of Eqs.~\eqref{eq:F_ab_sc} and \eqref{eq:F_ab_ann}
that the collision terms vanish when considering the interactions corresponding
only to sterile neutrinos.
When more than one sterile neutrino is considered, the damping terms between the different sterile neutrinos are therefore set to zero.

To summarise, the code computes the evolution of the
$N\times N$ neutrino density matrix
\begin{equation}\label{eq:varrho-B4}
\varrho(x, y)
=
\left(
\begin{array}{ccccc}
\varrho_{ee}&\varrho_{e\mu}&\varrho_{e\tau}&\varrho_{es_1}&\ldots\\
\varrho_{\mu e}&\varrho_{\mu\mu}&\varrho_{\mu\tau}&\varrho_{\mu s_1}&\\
\varrho_{\tau e}&\varrho_{\tau\mu}&\varrho_{\tau\tau}&\varrho_{\tau s_1}&\\
\varrho_{s_1e}&\varrho_{s_1\mu}&\varrho_{s_1\tau}&\varrho_{s_1s_1}&\\
\vdots&&&&\ddots
\end{array}
\right)\,,
\end{equation}
which is the same for neutrinos and antineutrinos,
and of the comoving photon temperature~$z$.
The momentum dependence of the density matrix $\varrho$ is taken into account
using a discrete grid of momenta, as described in section~\ref{ssec:momenta}.
The differential equations which the code solves are the following,
written in terms of the comoving coordinates
$x\equiv m_e\, a$, $y\equiv p\, a$ and $z\equiv T_\gamma\, a$
\cite{deSalas:2016ztq,Mirizzi:2012we,Saviano:2013ktj,Mangano:2001iu}:
\begin{eqnarray}\label{eq:drho_dx_nxn}
\frac{{\rm d}\varrho(y)}{{\rm d}x}
&=&
\sqrt{\frac{3 m^2_{\rm Pl}}{8\pi\rho}}
\left\{
    -i \frac{x^2}{m_e^3}
    \left[
        \frac{\mathbb{M}_{\rm F}}{2y}
        -
        \frac{8\sqrt{2}G_{\rm F} y m_e^6}{3x^6}
        \left(
            \frac{\mathbb{E}_\ell}{m_W^2}
            +
            \frac{\mathbb{E}_\nu}{m_Z^2}
        \right),
    \varrho
    \right]
    +\frac{m_e^3}{x^4}\mathcal{I(\varrho)}
\right\}\,,
\nonumber\\
\label{eq:dz_dx_nxn}
\frac{\mathrm{d}z}{\mathrm{d}x}
&=&
\cfrac{
{\displaystyle \sum_{\ell=e,\mu}}
\left[
\cfrac{r_\ell^2}{r} J(r_\ell)
\right]
+ G_1(r)
- \cfrac{1}{2\pi^2z^3}
    {\displaystyle \int_0^\infty \mathrm{d}y\,y^3\sum_{\alpha=e}^{s_{N_s}^{}}\cfrac{\mathrm{d}\varrho_{\alpha \alpha}}{\mathrm{d}x}}
}{
{\displaystyle \sum_{\ell=e,\mu}}
\left[
r^2_\ell J(r_\ell)
+ Y(r_\ell)
\right]
+ G_2(r)
+ \cfrac{2\pi^2}{15}
}\,,
\end{eqnarray}
where $r=x/z$ and $r_\ell=m_\ell/m_e\,r$.
The expressions for the $J$, $Y$, $G_1$ and $G_2$ functions,
which take into account the electromagnetic corrections to electron and photon masses,
are written in Eq.(18)--(22) of \cite{Mangano:2001iu}.
We report them here for completeness:
\begin{eqnarray}
J(r)
&=&
\frac{1}{\pi^2}
\int_0^\infty {\rm d}u \, u^2
\frac{\exp(\sqrt{u^2+r^2})}{\left[\exp(\sqrt{u^2+r^2})+1\right]^2}
\label{eq:j}
\,,\\
Y(r)
&=&
\frac{1}{\pi^2}
\int_0^\infty \mathrm{d}u \, u^4
\frac{\exp(\sqrt{u^2+r^2})}{\left[\exp(\sqrt{u^2+r^2})+1\right]^2}
\label{eq:y}
\,,\\
K(r)
&=&
\frac{1}{\pi^2}
\int_0^\infty {\rm d}u \, 
\frac{u^2}{\sqrt{u^2+r^2}}
\frac{1}{\exp(\sqrt{u^2+r^2})+1}
\label{eq:k}
\,,\\
G_1(r)
&=&
2\pi\alpha
\left[
  \frac{1}{r}
  \left(
    \frac{K}{3}
    + 2 K^2
    -\frac{J}{6}
    -KJ
  \right)
  +
  G_3
\right]
\label{eq:g1}
\,,\\
G_2(r)
&=&
-8\pi\alpha
\left(
  \frac{K}{6}
  +\frac{J}{6}
  -\frac{1}{2}K^2
  +KJ
\right)
+
2\pi\alpha r
G_3
\label{eq:g2}
\,,\\
G_3(r)
&=&
\frac{K'}{6}
-KK'
+\frac{J'}{6}
+K'J
+KJ'
\,,
\label{eq:g3}
\end{eqnarray}
where the prime denotes derivative with respect to $r$ and we dropped the explicit dependence on $r$ in the expressions for the $G$ functions.
For the sake of computational speed, we calculate and store lists for all the terms of Eq.~\eqref{eq:dz_dx_nxn}
which do not depend on the neutrino density matrix 
at the initialisation stage, and compute their values through interpolation during the real calculation.
The same happens for the energy densities of charged leptons, for which performing an interpolation
is much faster than computing an integral.

Finally, in order to estimate the effective comoving neutrino temperature $w\equiv T_\nu\,a$, which is
not needed for the calculation but useful to understand the final results,
we use an equation similar to \eqref{eq:dz_dx_nxn},
but considering only relativistic electrons, i.e.\ fixing $r_e=0$ in Eq.~\eqref{eq:dz_dx_nxn}.

\subsection{Solver and initial conditions}
\label{ssec:solver}
We solve the differential equations with the \dlsoda\ routine
from the \texttt{ODEPACK}%
\footnote{\url{https://computation.llnl.gov/casc/odepack/odepack_home.html}.}\
Fortran package \cite{hindmarsh1982odepack,dlsoda1}.
\texttt{ODEPACK} is a collection of solvers for the initial value problem for systems of ordinary differential equations.
It includes methods to deal with stiff and non-stiff systems, and some of the provided subroutines
automatically recognise which type of problem they are facing.

The specific solver we use, \dlsoda,
is a modification of the Double-precision Livermore Solver for Ordinary Differential Equations (\texttt{DLSODE})
which includes an automatic switching between stiff and non-stiff problems
of the form ${\rm d}y/{\rm d}t = f(t,y)$.
In the stiff case, it treats the Jacobian matrix ${\rm d}f/{\rm d}y$ as either a dense (full) or a banded matrix, and as either user-supplied or internally approximated by difference quotients.
It uses Adams methods (predictor-corrector) in the non-stiff case, and Backward Differentiation Formula (BDF) methods (the Gear methods) in the stiff case.
The linear systems that arise are solved by direct methods (LU factor/solve).
For more details, see the original publications  \cite{hindmarsh1982odepack,dlsoda1}.

The initial conditions for \dlsoda\ are defined as follows.
The initial time $x_{\rm in}$ is an input parameter of the code,
and reasonable values would correspond to temperatures between a few hundreds and a few tens of MeV.
The initial comoving photon temperature is computed evolving Eq.~\eqref{eq:dz_dx_nxn}
from even earlier times ($z_0=1$ at $T_0=10\, m_\mu$, $x_0=m_e/T_0$) until $x_{\rm in}$.
The obtained value $z_{\rm in}$ is then considered as the temperature of equilibrium
of the entire plasma. Concerning the neutrino density matrix at $x_{\rm in}$, all off-diagonal elements and the diagonal ones for sterile
neutrinos are fixed to zero, while the diagonal elements corresponding to active neutrinos are 
Fermi-Dirac distributions with a temperature $z_{\rm in}$.
For typical values that we use in the code,
we have $z_{\rm in}-1=2.9\e{-4}$ for $x_{\rm in}=0.001$ (which we use for the 3+1 cases)
or
$z_{\rm in} = 1.098$ for $x_{\rm in}=0.05$ (suitable for the three-neutrino case, see \cite{deSalas:2016ztq}).

\subsection{Momentum grid}
\label{ssec:momenta}
In order to follow the evolution of Eq.~\eqref{eq:varrho-B4}, we discretise its dependence on $y$ and evolve each of the momentum in $x$.
One of the most interesting ways to make the code more precise and faster is related to the choice of the $y_i$.
Discretising the momenta with a linear or logarithmic spacing works, but it is not the most efficient
way to generate the grid.
Inspired by one of the methods used in \texttt{CLASS} (see \cite{Lesgourgues:2011rh}),
we deeply tested and finally considered a spacing based on the Gauss-Laguerre integration method.
The crucial point of the calculation is to compute the energy density of neutrinos,
given by
\begin{equation}
\rho_\alpha
=
\frac{1}{\pi^2}
\int_0^\infty
\mathrm{d}y\, y^3\,\varrho_{\alpha\alpha}(y),
\end{equation}
where $\varrho_{\alpha\alpha}(y)$ will be close to a Fermi-Dirac distribution and in any case always exponentially suppressed.
The Gauss-Laguerre quadrature (see e.g.~\cite{NR})
is a method that is designed to optimise the solution of integrals of the type
\begin{equation}
I
=
\int_0^\infty
{\rm d}x\, y^\alpha\,e^{-y}\,f(y)
\simeq\sum_i^{N}w_i^{(\alpha)}\,f(y_i)
\,,
\end{equation}
where $f(y)$ is a generic function,
$y_i$ are the $N$ roots of the Laguerre polynomial $L_N$ of order $N$, and $w_i$ are relative weights,
 which are obtained using
\begin{equation}
w_i^{(\alpha)}=
\frac{y_i}{(N+1)^2[L^{(\alpha)}_{N+1}(y_i)]^2}.
\end{equation}
The weights can be computed for example using the \texttt{gaulag} routine from \cite{NR}.
Since our momentum distribution function is not directly proportional to $e^{-y}$,
we consider $f(x)=e^y\,\varrho_{\alpha\alpha}(y)$,
in order to rescale the weights appropriately.

For the simple purpose of integrating the Fermi-Dirac distribution, very few points are typically required.
\texttt{CLASS}, for example, uses order of ten points for integrating the neutrino distribution.
In our case the non-thermal distortions must be computed accurately, and in particular
when evolving the thermalisation of a sterile neutrino we need more precision on the small momenta.
On the other hand, we do not want to compute the momentum distribution function at very high $y$,
which gives a very small contribution to the total integral. We therefore use a truncated list of nodes $y_i$ over which to compute the evolution of $\varrho$,
selecting only the $N_y\leq N$ nodes for which $y_i<20$.
In this way we can increase the number of points at small $y$ and the resolution on the thermalisation processes
without having to compute a large number of points at high momentum.
The number of points we can use is limited by the accuracy of the algorithm that computes the $w_i$.
For the \texttt{gaulag} routine \cite{NR}, our setup allows to reach $N_y\sim50$ when $N\sim350$.
This number of momentum nodes is already large enough to reach a precision
much better than one per mille on \Neff, which is the same we could obtain with a linear spacing of the points
and $N_y=100$ \cite{deSalas:2016ztq}.
Since the evaluation of the collision integrals scales as $N_y^2$
and the number of derivatives in Eq.~\eqref{eq:drho_dx_nxn} scales with $N_y$,
this ensures a significant gain.
We further comment on this point in the next sections.

\subsection{Numerical calculation of 1D and 2D integrals}
\label{ssec:integrals}
Most of the processing time is spent to compute the collision integrals
discussed in section~\ref{sec:collint}, which are two-dimensional integrals in the momentum.
We compute the integrals using a two-dimensional version of the Gauss-Laguerre method,
which has been tested to be precise enough,
\begin{equation}
\int_{x_1}^{x_N}\int_{y_1}^{y_M}{\rm d}x\, \mathrm{d}y\, f(x, y)
=
\sum_{i=1}^{N}\sum_{j=1}^{M}
w_i\,w_j\,f_{ij}
\,.
\end{equation}
This works under the assumption that $f(x,y)$ is exponentially suppressed both in $x$ and $y$.
Such assumption is valid in our case, as the functions $F_{ab}$ always contain products of momentum distribution functions,
which are typically very close to the Fermi-Dirac.
The only exception is the case of the additional neutrino, for which the distribution can be very different from the Fermi-Dirac,
but in any case it is always exponentially suppressed, since the lowest momenta
are always populated first and its momentum distribution can never exceed the one of standard neutrinos.

When using a linear/logarithmic spacing of points, instead
we perform the integrals using a composite two-dimensional Newton-Cotes formula of order 1 \cite{newtoncotes}:
\begin{equation}
\int_{x_1}^{x_N}\int_{y_1}^{y_M}{\rm d}x\, {\rm d}y\, f(x, y)
=
\sum_{i=1}^{N-1}\sum_{j=1}^{M-1}
(x_j-x_i)(y_j-y_i)
\left[\frac{f_{ij} + f_{i+1,j} + f_{i,j+1} + f_{i+1,j+1}}{4}\right]\,,
\end{equation}
where we used the short notation $f_{i,j} = f(x_i,y_j)$,
while $i$ and $j$ run over the grid of momenta we are using,
which contains $N=M=N_y$ points for each dimension.
This avoids us the need to interpolate the density matrix in points outside the momentum grid.

The integrals therefore require $N_y^2$ evaluations of the integrands at each evaluation:
this means that reducing the value of $N_y$ by a factor of two
gives a factor four faster calculation of the integrals.
The actual gain in the code is even larger, since the \texttt{DLSODA} algorithm needs to explore less combinations
of variations in the $\varrho_{\alpha\beta}(y_l)$ for the different $y_l$ in the momentum grid.
Our goal is therefore to obtain with a coarse grid
a result that is in reasonable agreement with the one obtained using a fine grid.

In order to obtain the maximum speed,
we study the accuracy of each function that enters the code in comparison
with the analytical results, were they can be obtained.
The number of points and the integration methods adopted in all the integrals,
for example, have been carefully studied to achive a reasonable precision
with a short computation time.
For the two-dimensional integrals, the selected momentum grid
fully defines the integration procedure, and the precision is always good
when using a reasonable number of points.
Depending on the function, we may adopt the Gauss-Laguerre, Newton-Cotes or Romberg integration \cite{Romberg:1955}
methods for the one-dimensional integrals.
In particular, for the electron and muon energy densities
and for most of the funcions that enter the calculation of Eq.~\eqref{eq:dz_dx_nxn}
we use a Gauss-Laguerre method on a dedicated grid of up to 110 points
for the most complicated functions.
In one single case, the $K'(r)$ function derived from Eq.~\eqref{eq:k},
the result obtained with the Gauss-Laguerre method
did not reach the requested precision and we decided to use a Romberg integration instead.
Although this requires a longer computation time, it only affects the initialisation stage,
as in the code we interpolate over the pre-computed values.
The number of points and the interpolation range have also been studied in order to obtain sufficiently precise results
for all the computations required in the code.

\subsection{Precision of the final results}
\label{ssec:precision}
We have tested our code with the results available in the literature and
verified the robustness of our findings against changes in the settings used in the calculations.
In particular, we refer to the high-precision results in the three-neutrino case of \cite{deSalas:2016ztq},
from which we have adopted all the equations.

Concerning the value of \Neff\ that we obtain using only active neutrinos,
we verified that we can reach much better than per mille stability on $\Neff=3.044$
using $N_y\geq20$ points spaced with the Gauss-Laguerre method,
if the tolerance for \dlsoda%
\footnote{For simplicity, we assume the same numerical value for both the relative and absolute tolerance.
The algorithm will always match the most stringent of the two requirements.}
is $10^{-6}$.
This means that using $N_y=50$ instead of $N_y=20$ does not significantly alter the result.
If we want to consider a linear or logarithmic spacing for the momentum grid,
a minimum of 40 grid points must be employed in order to reach the same level of stability.
Another possible setting that can give us a faster execution of the code is the precision
used for \dlsoda.
We verified that once the tolerance for \dlsoda\ is smaller than $10^{-5}$,
the results are already stable at a level much better than per mille
(actually closer to the 0.1 per mille)
with respect to the most precise case considered here
($N_y=50$, tolerance $10^{-6}$).
Using a tolerance of $10^{-4}$ gives a value of \Neff\ which is stable
at the level of few per mille, and still better than 1\%.

If we repeat the same exercise in the 3+1 scheme,
using $\dmsq{41}=1.29$~eV$^2$,
$\uasq{e}=0.012$ \cite{Gariazzo:2018mwd} and $\uasq{\mu}=\uasq{\tau}=0$,
we find similar conclusions.
A tolerance of $10^{-5}$ gives results very close to those obtained with $10^{-6}$,
while any larger tolerance gives larger fluctuations depending on $N_y$.
With $10^{-4}$, the precision remains of the order of 0.5\%, so it is still safe to compute the value of \Neff\ on a grid of active-sterile mixing parameters
using this level of precision.
With $N_y=20$, a single run takes a few minutes on four cores, and the running time is not significantly affected
by changes in the \dlsoda\ tolerance.
When more precision is required, however, the algorithm may have troubles in resolving some of the resonances,
and in that case the run can take much longer because of the adaptive nature of the solver.

Another parameter that we tested is the initial value of $x$, $x_{\rm in}$.
Apart for fluctuations which are compatible with those obtained varying $N_y$,
the result is stable against variations in $5\e{-4}\leq x_{\rm in}\leq5\e{-2}$.
The largest values of $x_{\rm in}$ may be inappropriate for high values of \dmsq{41},
as it is important for the solver to start the evolution before the sterile state
starts to oscillate significantly with the active ones.
Smaller values, on the contrary, may create numerical problems in \dlsoda\
due to the very small initial value $z_{\rm in}-1$,
and are never really required for our purposes.

\bibliography{main}

\end{document}